\documentclass[useAMS,usenatbib]{mn2e}

\usepackage{graphicx}

\title[Decomposition of Barred Galaxies and AGN]{Image Decomposition of Barred Galaxies and AGN Hosts}
\author[Dimitri Gadotti]{Dimitri Alexei Gadotti\thanks{E-mail: dimitri@mpa-garching.mpg.de}\\
Max-Planck-Institut f\"ur Astrophysik, Karl-Schwarzschild-Str. 1, D-85748
Garching bei M\"unchen, Germany}
\begin{document}


\pagerange{\pageref{firstpage}--\pageref{lastpage}} \pubyear{2007}

\maketitle

\label{firstpage}

\begin{abstract}
I present the results of multi-component decomposition of V and R broadband images of a sample
of 17 nearby galaxies, most of them hosting bars and active galactic nuclei (AGN). I use {\sc budda} v2.1
to produce the fits, allowing to include bars and AGN in the models. A comparison with previous results
from the literature shows a fairly good agreement. It is found that the axial ratio of bars, as measured
from ellipse fits, can be severely underestimated if the galaxy axisymmetric component is relatively luminous.
Thus, reliable bar axial ratios can only be determined by taking into account the contributions of bulge
and disc to the light distribution in the galaxy image. Through a number of tests, I show that
neglecting bars when modelling barred galaxies can result in a overestimation of the bulge-to-total
luminosity ratio of a factor of two. Similar effects result when bright, type 1 AGN are not considered
in the models. By artificially redshifting the images, I show that the structural parameters of more distant
galaxies can in general be reliably retrieved through image fitting, at least up to the point where the
physical spatial resolution is $\approx$ 1.5 Kpc. This corresponds, for instance,
to images of galaxies at $z=0.05$ with a seeing FWHM of 1.5\arcsec, typical of the SDSS. In addition, such
a resolution is also similar to what can be achieved with HST, and ground-based telescopes with adaptive
optics, at $z\sim1-2$. Thus, these results also concern
deeper studies such as COSMOS and SINS. This exercise shows that
disc parameters are particularly robust, but bulge parameters are prone to errors if its effective
radius is small compared to the seeing radius, and might suffer
from systematic effects. For instance, the bulge-to-total luminosity ratio is systematically overestimated,
on average, by 0.05 (i.e., 5\% of the galaxy total luminosity). In this low resolution
regime, the effects of ignoring bars are still present, but AGN light is smeared out. I briefly discuss the
consequences of these results to studies of the structural properties of galaxies, in particular on
the stellar mass budget in the local universe. With reasonable assumptions, it is possible to show that the
stellar content in bars can be similar to that in classical
bulges and elliptical galaxies. Finally, I revisit the cases of NGC 4608 and NGC 5701
and show that the lack of stars in the disc region inside the bar radius is significant. Accordingly,
the best fit model for the former uses a Freeman type II disc.
\end{abstract}

\begin{keywords}
galaxies: bulges -- galaxies: evolution -- galaxies: formation -- galaxies: fundamental parameters --
galaxies: photometry -- galaxies: structure
\end{keywords}

\section{Introduction}

Parametric modelling of galaxy images has recently become a popular tool to measure structural parameters,
such as scale-lengths and stellar masses, of the different galactic components, particularly bulges and discs.
Through this sort of analysis, one is also able to determine the relative importance of the bulge component,
with parameters such as the bulge-to-total luminosity ratio B/T, one of the major attributes that
define the Hubble sequence \citep{hub26,hub36}. It thus provides indispensable means to investigate the formation
and evolution of galaxies, and the origin of the Hubble sequence, some of the key subjects in current
astrophysical research. Such studies can be divided in two categories. In the first
category, one usually finds samples of some tens of very nearby ($z<0.01$) galaxies \citep[e.g.,][]{dej95,
kho00,mol01,don01,pen02,des04,lau04,lau05,lau06}. In this case, it is possible to fit models on a more careful,
individual basis, and study other components, such as bars, lenses, rings and spirals, either by including
them in the models or analysing residual images, where the fitted model is removed from the original galaxy image.
In the second category, one usually finds samples of some hundreds or thousands of galaxies up to $z\sim1$
\citep[e.g.,][see also \citealt{pig06}]{mar98,tas05,all06,hau07,hue07}, where fits are done in an automated fashion
and structural details are ignored, often smoothed out by the low physical spatial resolution. In this case,
although individual fits might not be fully reliable, solid statistical analyses are attainable.

Studies in both categories provide observational constraints to test theoretical predictions from models
or scenarios of galaxy formation and evolution. Consequently, in order not to hamper progress, it is critical
to understand the weaknesses and biases of these techniques. In fact, image decomposition of galaxies is
a complex and difficult endeavour. One has to define which components to fit, the model(s) to adopt, which code
to use, or basically the algorithm in the search for the best fit, and the initial setup for the
fitting process. Furthermore, there are issues related to seeing effects, sky subtraction, crowding,
spatial resolution and signal-to-noise ratio. Each one of these points affects the resulting physical
parameters to some extent.

In this paper, I make a connection between the two categories and study some of the issues that might
produce wrong results. First, I present the results of image decomposition of a sample
of galaxies at $z\sim0.005$. Most of these galaxies are barred and host AGN, and these components
are also fitted in the models. This provides structural parameters of bulges, bars and discs, and
is particularly relevant, since it is in the study of galaxies with non-axisymmetric components like bars
where 2D, image fitting is most advantageous over 1D, luminosity profile fitting. In addition, bars are known
now to play a major role in galaxy evolution \citep[e.g.,][and references therein]{kor04} and thus a better
understanding of this galactic component is needed. Then, I remove bars
and AGN from the models and redo the fits to verify how harmful it may be to one's results not to include these
components in the model when they are clearly present in the galaxy. This is motivated by the fact that
most studies ignore bars, focusing on estimating the structural parameters of bulges and discs, even
though bars are ubiquitous and can represent a significant fraction of a galaxy luminosity. In fact,
\citet{esk00} find that only 27\% of a sample of 186 spiral galaxies is unbarred in the near-IR, and
the fraction of the total luminosity of a galaxy in the bar can be as high as $\sim30\%$
\citep[see][]{sel93}. Furthermore, results in \citet{lau05} and \citet{lau06} hint that light from the bar can,
at least in some cases, be attributed to the bulge if the bar is omitted in the model. In principle, this
can also happen to the light from a bright AGN. Finally, I use artificially redshifted images of the galaxies
in the sample to redo the fits and repeat the exercise of omitting bars and AGN. This allows me to check
the effects of low resolution on the fits and the importance of having more detailed models when studying
more distant galaxies.

This paper is organised as follows. In the next section, I briefly recall the relevant aspects of the data
sample used here, referring the reader to \citet[][hereafter GdS06]{gad06},
where one also finds a more detailed presentation of the acquisition and treatment
of the raw data. In Sect. 3, I present the results from image decomposition with and without bars and AGN
in the models, including a comparison with previous work. The corresponding results for the redshifted images
are presented in Sect. 4. The implications of this work to studies based on image decomposition and
structural analysis of galaxies are discussed in Sect. 5. I summarise and present the main conclusions in
Sect. 6.

\section{The Data}

The sample consists of 17 disc galaxies selected in GdS06 at an average redshift of about 0.005. Basic data
for these galaxies are displayed in Table \ref{basic}. They are relatively bright,
most are close to face-on, and have no strong morphological perturbations. These characteristics usually
ensure a reliable structural analysis. This sample was selected to cover the various relevant properties
of barred galaxies with morphological type earlier than Sbc.
Only two galaxies have no clear bar in their images, although one of
them is classified as weakly barred in the RC3 \citep{dev91}, and the others
have bars which range from very weak to very strong features. For five galaxies (all classified as Seyferts)
an AGN component is deemed necessary in the fitted model, but most of the galaxies in the sample are classified
as having some type of AGN in the NASA Extragalactic Database (NED).
The positions in the Hubble sequence of the galaxies in the sample go from S0 to Sbc, according to the RC3.
Thus, although this sample is not in any sense complete or unbiased, it has appropriate qualities to study
the issues on image decomposition described in the Introduction, since it covers a suitable range
of bulge, bar and AGN properties. This will become clearer below, with the results from the structural analysis,
where one sees a relatively wide range of, e.g., bulge and bar prominence, AGN luminosity and bar
ellipticity and shape.

The images used in this work were obtained at the Kuiper 1.55 m telescope operated by the University of Arizona
Steward Observatory, on mount Bigelow. They were taken using Johnson-Morgan V and R broadband filters, with a
pixel size of 0.29\arcsec, and a square field of view of roughly 5\arcmin\ on a side. The total integration
time was 1500 s in V and 900 s in R. The seeing FWHM was about 1.3\arcsec\ on average. Errors in the photometric
calibrations are $\approx$ 0.02 mag, but not all nights were photometric. One could argue that a larger data
set could be obtained, for instance, from the SDSS. It should be noted, however, that the images collected
in GdS06 have a much higher signal-to-noise ratio, and were, on average, obtained under better seeing.
In addition, the relatively small sample provides an opportunity to carefully check every decomposition
performed, and assure that reliable results are obtained on an individual basis.

Further details on the properties of these galaxies, as well as on the observations and data reduction, can be
found in GdS06. It is important to note that, for some of the galaxies, since they cover practically
the whole field of view, a good estimate of the sky contribution was difficult to obtain. Likewise, in
some cases, seeing measurements are not very accurate, since there were only a few stars in the field of view
suitable for that. For these reasons, I excluded two galaxies from the original sample in
GdS06, as well as all the B and I broadband images, where the difficult sky subtraction might produce
errors in the image decomposition. The inaccuracy in the seeing measurements, though, might not be too severe,
as it was quite stable during the observing runs.

\section{Decomposition of Original Images}
\label{sec:decomps}

In this section, the results of image decomposition using the original images are presented. I will first
describe relevant technical aspects of the method applied, and then
show results using the full capabilities of the image decomposition code, i.e., its ability to model
bars and AGN when necessary. Afterwards, I will show, in two separate subsections, the effects
of not including these components in the models, when fitting galaxies which clearly have them.

\subsection{Method and Fitting Functions}

The code used here for the decompositions is {\sc budda} v2.1 \citep[see][]{des04}. This latest version
of the code has several improvements, as compared to its first version. Beside models for bulge and disc,
there is now the possibility of fitting bars and a central source of light, like an AGN. The PSF is now
described by a \citet{mof69} profile, which has been shown to reproduce the effects of atmospheric turbulence
better than a Gaussian profile \citep{tru01b}. An improvement was also made to the way errors are computed for
each parameter estimated by the code. The errors are estimated after the convergence of the code to the
global $\chi^2$ minimum. Each parameter is varied successively until the new $\chi^2$
reaches a threshold equivalent to 1 $\sigma$ for a normal $\chi^2$ probability distribution.
This often produces meaningless error values if the parameter under consideration does not influence
significantly the final $\chi^2$ value of the model. For instance, if one has a faint disc in a lenticular
galaxy, changing the disc parameters by a large amount will only produce small variations in the
$\chi^2$ of the total model. The errors obtained for the disc parameters will then be very large and
meaningless. The same happens for the position angle and ellipticity of a component that is close to
circular. To obtain useful error estimates, the code now weights the $\chi^2$ threshold of each
parameter by the fraction of light from the total model that comes from the component to which the
parameter refers. The $\chi^2$ threshold of the geometric parameters (position angle and ellipticity) are
weighted by the ellipticity of the component. All new features in the code were tested both with synthetic and
with real galaxy images. In particular, tests with artificial galaxies demonstrated that the parameters
retrieved are often within 1 $\sigma$ of the real, input value, and almost always within 3 $\sigma$. This
indicates that the new procedure for error estimation gives reliable results. Nevertheless, it should be
noted that these errors are purely statistical, do not take into account other sources of
uncertainty, and thus must be considered as lower limits of the true error.

The disc surface brightness profile is described by an exponential function \citep[type I disc,][]{fre70},

\begin{equation}
\mu_d(r)=\mu_0 + 1.086r/h,
\end{equation}

\noindent where $r$ is the galactocentric distance, $\mu_0$ is the disc central surface brightness,
and $h$ is the disc characteristic scale-length.

The bulge surface brightness profile is described by a S\'ersic function
\citep[][see \citealt{cao93}]{ser68},

\begin{equation}
\mu_b(r)=\mu_e+c_n\left[\left(\frac{r}{r_e}\right)^{1/n}-1\right],
\end{equation}

\noindent where $r_e$ is the effective radius of the bulge, i.e., the radius that contains half of its
light, $\mu_e$ is the bulge effective surface brightness, i.e., the surface brightness at $r_e$, $n$ is the
S\'ersic index, defining the shape of the profile, and $c_n=2.5(0.868n-0.142)$.

The AGN is modelled as an unresolved point source convolved with the PSF Moffat profile. The FWHM of the AGN profile
has thus the same value of the seeing, and the only parameter fitted by the code is its peak intensity.

The bar luminosity profile is also described by a S\'ersic function. For the bar,

\begin{equation}
\mu_{\rm Bar}(r)=\mu_{e,{\rm Bar}}+c_{n,{\rm Bar}}\left[\left(\frac{r}{r_{e,{\rm Bar}}}\right)^{1/n_{{\rm Bar}}}-1\right],
\end{equation}

\noindent where $c_{n,{\rm Bar}}=2.5(0.868n_{{\rm Bar}}-0.142)$, and the other parameters have similar definitions
as for the bulge. Another bar parameter fitted by the code is the length of the bar semi-major axis, $L_{{\rm bar}}$,
after which the bar light profile is simply truncated and drops to zero.

Except for the AGN, which is circular, the model components are described by concentric, generalised ellipses
\citep[see][]{ath90}:

\begin{equation}
\left(\frac{|x|}{a}\right)^c+\left(\frac{|y|}{b}\right)^c=1,
\end{equation}

\noindent where $x$ and $y$ are the pixel coordinates of the ellipse points, $a$ and $b$ are the extent of its
semi-major and semi-minor axes, respectively, and $c$ is a shape parameter. Position angles and ellipticities
($\epsilon=1-b/a$) were fitted by the code for every component. When $c=2$ one has a simple ellipse.
When $c<2$ the ellipse is discy, and when $c>2$ the ellipse is boxy. For bulges and discs I fixed $c=2$ but
this parameter was left free to fit bars, since these components are better described by boxy ellipses.

\citet{lau05} used a function to describe the bar luminosity profile which corresponds to a projected surface density
of a prolate Ferrers bar \citep[see][]{bin87}, and showed that this results in good fits.
However, they also argued that using a S\'ersic function
is equivalent. In fact, the S\'ersic function proves to be very useful to describe the light distribution in bars. As
shown by \citet{elm85}, bars in late-type spirals generally have an exponential luminosity profile, whereas bars
in early-type spirals and lenticulars have a flatter luminosity profile. This duality can be conveniently
expressed with the S\'ersic index $n$: when $n\approx1$ the S\'ersic function is close to an exponential function, while
for $n<1$ one has a flatter profile. Hence, with a single function, it is possible to fit the different bar types, and
quantify this difference with a single parameter. The fact that we use the same function
for the luminosity profiles of bulges and
bars generally does not increase the possibility of a degenerate solution. This is mainly due to two reasons. Firstly,
the geometric properties of bulges and bars in galaxy images are markedly different: bulges are rounder and centrally
located while bars are more eccentric and extended. Secondly, the shape of their luminosity profiles, given by $n$ and
$n_{{\rm Bar}}$, is in most cases also different: from the results below, one sees that $n$ is
approximately in the range of 1 to 3, whereas $n_{{\rm Bar}}$ ranges from $\approx0.5$ to $\approx1$. These differences
give further constraints to find the best fitted model for each component. Note that when the S\'ersic index is
below $\approx0.15$ the luminosity profile has a depression in its central parts. However, for the luminosity
volume density, this occurs when the S\'ersic index is below 0.5 \citep{tru01a}.
We will see below that in a few cases the value found for $n_{{\rm Bar}}$ is between $\approx0.3$ and $\approx0.5$.
With the uncertainties in the determination of this parameter, one can not state firmly that these bars have outer
parts more luminous that their inner parts, only based on that. Nevertheless, given
the complex orbital structure of bars \citep[see, e.g.,][]{pat03}, it is not very surprising to find
bars with such property: it is well known that many bars do show bright structures at their ends, known as
ansae, and some such structures are found in the residual images below \citep[see, e.g.,][]{val07,but06}.

The initial setup of {\sc budda}, where rough estimates for each parameter are given as a starting
point to the code, was defined with visual inspection of the images and surface brightness radial profiles.
It should be noted that, in order to achieve the best fit, the code was run several times for each image,
in an interactive fashion, trying different initial setups, and checking the corresponding results through
comparisons between the galaxy and model surface brightness profiles, the $\chi^2$ value, estimated errors
and residual images. This is particularly relevant when there is doubt about what components to include
in the model. In this study, that did not happen often, but when it did it was most of the times concerning
the AGN component. In these cases, the code usually shows if the component is absent in the galaxy, indicating
values close to zero for its luminosity parameter. It is interesting to note that, even if a given galaxy
is classified as having an AGN, this component does not necessarily need to be taken into account in a
photometric model, since it can be an obscured, or a type 2, AGN, or simply not bright enough. This is in fact
the case for some of the galaxies in our sample.

\subsection{Results}

Figure \ref{results} shows, for each galaxy in the sample, the original image (at two different display levels,
as to emphasise either outer or inner parts), an image of the model obtained with {\sc budda}, and a residual image,
obtained after dividing the galaxy image by the model image, both in ADU. In the residual image, brighter shades
indicate regions where the model is more luminous than the galaxy, whereas darker shades indicate regions where
the model is fainter than the galaxy. Figure \ref{results} also shows surface brightness
radial profiles of the galaxy, of each component in the model separately, and of the total model, for comparison.
Two sorts of brightness profile are shown. One is obtained directly from the pixel values in the corresponding images,
following \citet{lau05}, and the other is obtained from ellipse fits, as usual. The results from ellipse fits also
include radial profiles of position angle, ellipticity and the b$_4$ Fourier component. The ellipse fits were done using
{\sc iraf}\footnote{{\sc iraf} is distributed by the National Optical Astronomy Observatories,
which are operated by the Association of Universities for Research in Astronomy, Inc., under
cooperative agreement with the National Science Foundation.} task {\sc ellipse}. The surface brightness values
are corrected for dust extinction in the Milky Way, and an inclination correction for intrinsic attenuation by
dust was also applied (see GdS06 for further details). Figure \ref{results} refers to our R-band images but
similar results were found in the V-band.
The values of the relevant structural parameters obtained are displayed in Tables
\ref{rres} and \ref{vres} for both bands. Note that the disc parameters obtained for NGC 4151, NGC 4665 and
NGC 5850 suffer from large uncertainties, due to their intrinsic low surface brightness and the issues on
sky subtraction mentioned above.

From Fig. \ref{results} it is possible to verify that the models obtained are a fairly good representation of their
corresponding galaxies, either checking the images provided or the radial profiles. Inspecting the residual images,
it is possible to identify many sub-structures, such as spiral arms (e.g., NGC 4394), inner discs (e.g., NGC 4151),
inner spirals (e.g., NGC 4314), nuclear rings (e.g., NGC 4593), inner bars (NGC 5850), ansae at the ends of the bar
(e.g., NGC 4151, NGC 4608, NGC 5850), dust lanes (e.g., NGC 4303 and NGC 5383)
or more complex dust structure (e.g., NGC 2110 and NGC 2911). Although some such structures can be
seen in the original galaxy images, in all cases they stand out much more clearly in the residual images.
This confirms residual images as a powerful tool to study these galactic components.
It is worth noting that, for many of our barred galaxies (but not all), the residual images reveal a very elongated
structure, inside the bar, along its major axis. This confirms the theoretical study of \citet{ath92},
where the morphology of orbits in barred galaxies is analysed. She found that, close to the major axis of the
bar, the dominant family of orbits is indeed very elongated, and that the orbits become less eccentric
away from the major axis (see her Fig. 12). It seems that the ellipsoid that fits the bar is able to account for
the external, less eccentric orbits, which are spread over most of the bar,
but the very elongated inner orbits show up in the residuals. It is plausible that a model
for the bar with an ellipticity that varies radially could fit most of the orbits. Alternatively,
one could use two ellipsoids for the bar: one as used here, and another, much more eccentric, but this is
beyond the scope of this study. It is also interesting to note that the same residual images show another
structure within the bar, but this is only in the central region of the galaxy. This could be associated
with an inner disc or a lens.

As mentioned, two types of surface brightness profile are shown. The one derived pixel by pixel has the
advantage of displaying information from the image as a whole, which is hidden in the elliptically
averaged profile from ellipse fits. The spread of the points in each pixel by pixel profile indicates
{\bf (i)}, in the models, the geometry of the corresponding model component (circular components have very narrow
profiles), and {\bf (ii)}, in the galaxies, their geometrical properties and the deviation from the average of the
light distribution through the galaxy surface, due to its own features (e.g., spiral arms, dust, star forming regions)
and statistical fluctuations in the photometry. Nonetheless, the usual ellipse fit profile is shown for the sake
of comparison. From these profiles, one sees models that range from excellent fits (e.g., NGC 2911, NGC 3227,
NGC 4267), corresponding usually to more simple galaxies, to fits where residuals are significant (e.g., NGC 4303,
NGC 4314, NGC 5850), usually galaxies with bright spiral arms and complex structure. Nevertheless, a typical difference
between galaxy and model is about only $\pm0.25$ mag arcsec$^{-2}$. It is interesting to note that, while bulges
and discs dominate the inner and outer parts of galaxies, respectively, bars can be the dominant structure at
intermediate radii (see IC 486, NGC 4314, NGC 4608, NGC 4665, NGC 5701 and NGC 5850).

Interestingly enough, from the ellipticity and
position angle profiles in Fig. \ref{results}, one sees that only by including a bar
it is possible to fit the rise and drop in ellipticity and the
abrupt changes seen in position angle, both features typical of barred galaxies
(see, e.g., NGC 4151, NGC 4267 and NGC 4608). The remaining
discrepancies in the position angle and ellipticity profiles seem to be caused by other components, such as spiral arms,
rings, and ovals \citep[see][]{gad07}, present in the galaxy but not in the models.
For instance, this is clear in NGC 4303, NGC 4314,
NGC 4394 and NGC 5701. In the models, after the bar end, usually accompanied by a sudden change in position angle
and a drop in ellipticity, these parameters assume the values of the disc component. In the galaxies, however,
if there is an additional component between the bar and the outer disc, with position angle and ellipticity different
from those of the disc, then there will be differences between the corresponding galaxy and model radial profiles.
The radial profiles of b$_4$ are the ones with strongest disagreement between galaxies and models. Either
the models do not reproduce the position where the peak in b$_4$ happens, or the corresponding b$_4$ values.
There is no clear reason why these discrepancies occur. Nevertheless, NGC 3227, NGC 4151 and NGC 5850, where
these disagreements are particularly severe, either show conspicuous dust structure or very bright ansae
at the end of the bar.

It is interesting to verify if known results can be reproduced with the output from the decompositions shown here.
Figure \ref{korm} shows the correlation between the effective surface brightness and the effective radius of bulges,
for the galaxies in the sample, in both bands. This correlation was first found by \citet{kor77} for elliptical
galaxies and afterwards shown to hold also for bulges. A similar correlation was found between the central surface
brightness and the scale-length of discs \citep[see, e.g.,][]{dej96}. Figure \ref{disks} shows that this correlation
is reproduced when one considers our V-band images, albeit with a large scatter. The scatter is even larger when one
looks at the results from our R-band images, rendering the correlation not significant in this case, in particular
if one ignores the outlying point, which corresponds to NGC 5850, whose disc parameters suffer from large uncertainties.
However, it is reasonable to assume that this is mostly a statistical effect, as the sample is not particularly large,
and thus that the correlation is real.
The length of bars is also correlated with the disc scale-length, as shown in Fig. \ref{bars} and by, e.g., \citet{erw05}.
This figure also shows that the bar lengths estimated with {\sc budda} agree very well with the
estimates from ellipse fits in GdS06. This is not surprising, since the latter are used as a starting point
for the code, but it is interesting to find such a good agreement as this parameter was not kept fixed in the
fits.

\subsection{Comparison with Previous Fits}
\label{sec:complau}

Figure \ref{complau} shows a comparison between some structural parameters obtained in this work with those found in
\citet{lau04} and \citet{lau05} for twelve galaxies also studied by them. In these papers, the structural parameters are
also obtained via an image decomposition code, able to fit bulges, discs and bars. The main methodological differences
between this study and theirs are: {\bf (i)}, they used a Ferrers model to describe the bar luminosity profile, {\bf (ii)},
they fitted an additional component (an oval) in NGC 4608, {\bf (iii)}, they have not modelled the AGN light
contribution, and {\bf (iv)}, they used near-infrared images. Figure \ref{complau} shows that there is generally good
agreement between our measurements. For 10 of the 12 galaxies, they have fixed the bar model to a very flat luminosity
profile, but it seems that the use of a different bar model does not lead to significantly different results.
The modelling of the oval in NGC 4608 also did not produce a fit significantly different from the one shown here. However,
I will show below that a better fit to this particular galaxy can be achieved using a Freeman type II disc \citep{fre70},
rather than a type I. Nevertheless, for some galaxies, the values of some structural parameters obtained here and
by Laurikainen et al. are discrepant. Most of these cases are pointed out in the figure. As already mentioned,
the fit to NGC 4151 is dubious. Likewise, Laurikainen et al. state that their results to NGC 3227, NGC 4151 and NGC 5701
can also be compromised: the authors mention that their images are not deep enough and the resulting parameters are
uncertain. Since their images are in the near-infrared and ours in the optical, and given that galaxies
usually become bluer outwards \citep[but see][]{gad01}, the scale-lengths obtained here could indeed be somewhat larger
than theirs. This systematic difference seems to be present in Fig. \ref{complau} but complicates these comparisons.
This difference in wavelength could also explain the different results for NGC 3227, NGC 4314 and NGC 4593, which have
significant amounts of dust and star formation which certainly have a stronger effect in the optical than in the
near-infrared. In particular, NGC 4314 has very bright star-forming nuclear spiral arms that show up clearly in the
residual image. The light from these spirals is partially
attributed to the bulge component, and, since these spirals should
not be so conspicuous in the near-infrared, this can explain why I obtain a much more massive bulge than
Laurikainen et al. On the other hand, it is not clear why our results are discrepant for the bulge of NGC 5701,
although the difference in the S\'ersic index is within typical 1 $\sigma$ errors.
Furthermore, it is unclear if the AGN light should have been taken into account also in their
near-infrared images. Taken altogether, this general agreement, and the fact that with the results presented here
it is possible to reproduce some known results (Figs. \ref{korm}, \ref{disks} and \ref{bars}) are very encouraging.

\subsection{The Ellipticity of Bars}

To measure the ellipticity of a bar, one usually fit ellipses to the galaxy image and assume that the
bar ellipticity is that of the most eccentric fitted ellipse \citep[e.g.,][]{mar07}. This is a very important
parameter because it is strongly related with the strength of the bar, and can provide constraints for bar models, like
those of, e.g., \citet{ath02} and secular evolution scenarios \citep[e.g.,][]{gad01}. In Fig. \ref{ecc}, the ellipticity
of bars as estimated from the image decompositions are plotted against the ellipticity peak in the ellipse fits of GdS06.
The latter is used as a starting point for the decompositions but the bar ellipticity is a free parameter in the fits.
It is clear that ellipse fits underestimate the true ellipticity of the bar. This effect is on average about 20\%, but
it can be as large as a factor of three. For NGC 5850, however, the match between the two parameters is excellent.
Examining the cases where this effect is strongest, as in NGC 4267 and NGC 4303, reveals its origin: the ellipticity of
the isophotes in the bar region is diluted by the contribution from the round, axisymmetric component of the galaxy.
The strength of this effect is thus governed by the difference between the contributions of the bar and the disc to the
total light in the galaxy near the bar end, since it is about this region where the bar isophotes reach their peak
in ellipticity, and the bulge component is usually faint there. As the disc of NGC 5850 is very faint, this dilution is
not efficient in this galaxy.

These results have implications on previous findings in the literature. For instance, \citet{mar07} measured
the bar ellipticity via ellipse fits in a sample of 180 barred galaxies. They found that only a minority of
the bars in their sample have ellipticities below 0.4, and that most bars have ellipticities between 0.5 and 0.8, with
a mean value of about 0.5. As I have just shown, ellipse fits systematically underestimate the true bar ellipticity
by 20\%, on average. This means that the paucity of weak bars, i.e., those with ellipticities below 0.4, is even
more pronounced. Many of these bars in their sample might have higher ellipticities, which, however, can not be
reliably measured with ellipse fits, due to the effects just discussed. Furthermore, the fraction of bars with
high ellipticities is, in fact, even higher, as is the true mean value for the ellipticity of bars. A simple
calculation gives a true mean value of around 0.6, considering an underestimation of 20\%.

It should be noted that the bar strength does not depend only on the bar ellipticity, but also on its shape (more
rectangular, boxy bars, i.e., those with higher $c$, are stronger) and its mass \citep[see also discussion in]
[and references therein]{mar07}.
The strength of a bar is thus directly connected to the amplitude of the non-axisymmetric potential it introduces
in the overall potential well of its host galaxy. With this definition, the strength of a bar does
not depend on any other of the properties of its host galaxy. On the other hand, the {\em impact} of a bar on the
evolution of a galaxy depends on other galaxy properties. A strong bar will significantly modify
the dynamics of gas and stars in a galaxy with a relatively weak axisymmetric potential, i.e., a galaxy where
the bar mass is high compared to the bulge and disc masses. The same bar would produce less significant effects
in a galaxy with massive bulge and disc. Furthermore, since the axisymmetric component of the potential is
centrally concentrated, mainly due to the bulge, the changes due to the bar are likely to have a dependence
on radius and to be more significant closer to the bar ends.

\subsection{Disc Consumption in NGC 4608 and NGC 5701}
\label{4608}

\begin{figure*}
   \centering
   \includegraphics[height=4cm,clip=true]{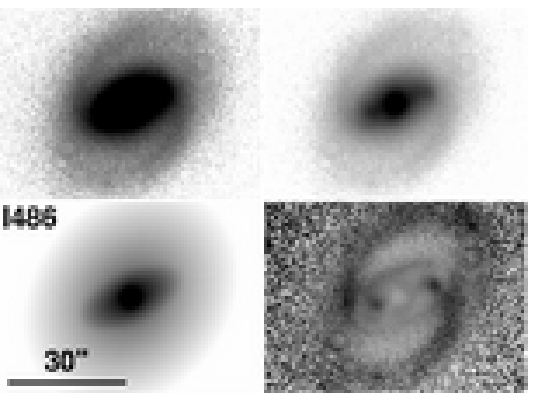}
   \includegraphics[height=4cm,clip=true]{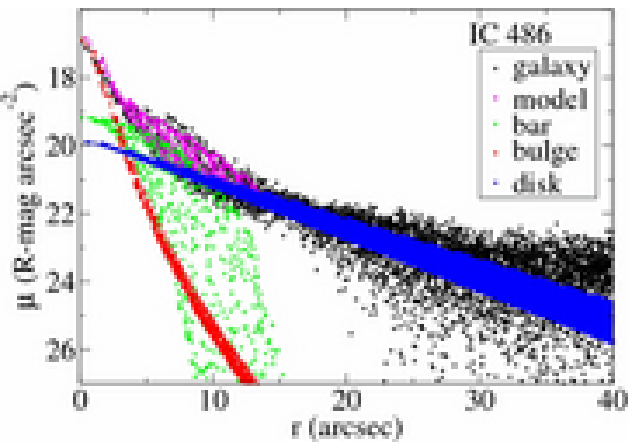}
   \includegraphics[height=4cm,clip=true]{f1I486c.eps}
   \includegraphics[height=4cm,clip=true]{f1I486d.eps}
   \caption{Results of image decomposition in the R-band for each galaxy in the sample. The images on the left
show the galaxy with emphasis on its outer (top left) and inner parts (top right), as well as the model
and residual images (bottom left and right, respectively). In the latter, brighter shades indicate regions where
the model is more luminous than the galaxy, whereas darker shades indicate regions where the model is fainter
than the galaxy. The panel at the centre shows the surface brightness
profiles of the galaxy and the models as indicated. Each point in these profiles correspond to a single pixel.
Only 10\% of the pixels are shown. The panels on the right show the results of ellipse fits to the galaxy and
model images. These are radial profiles of surface brightness (elliptically averaged) with residuals, and
geometric parameters: position angle (from North to East -- top), ellipticity (centre) and the b$_4$ Fourier
coefficient (bottom).}
\label{results}
\end{figure*}

\begin{figure}
   \centering
   \includegraphics[width=5cm,clip=true]{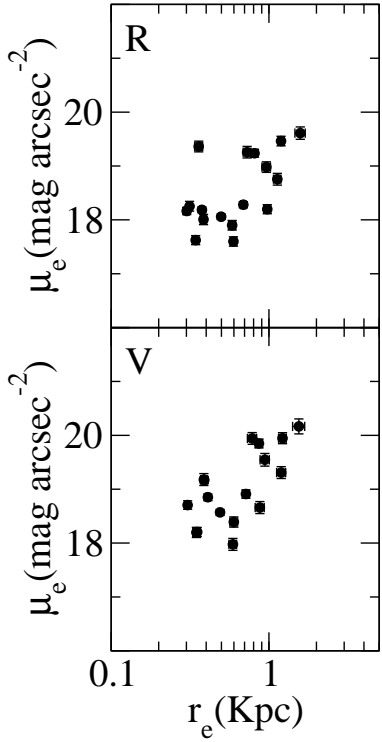}
   \caption{Correlation between the effective surface brightness and the effective radius of bulges, for the galaxies in
the sample, in both bands.}
   \label{korm}
\end{figure}

\begin{figure}
   \centering
   \includegraphics[width=8.4cm,clip=true]{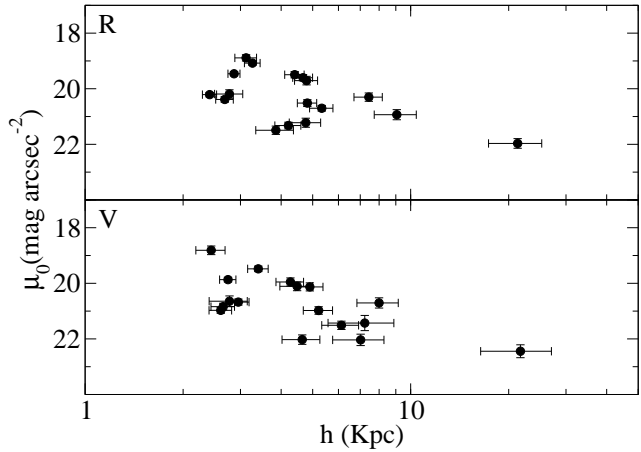}
   \caption{The central surface brightness of discs plotted against their scale-lengths, for the galaxies in
the sample, in both bands, as indicated.}
   \label{disks}
\end{figure}

\begin{figure}
   \centering
   \includegraphics[width=8.4cm,clip=true]{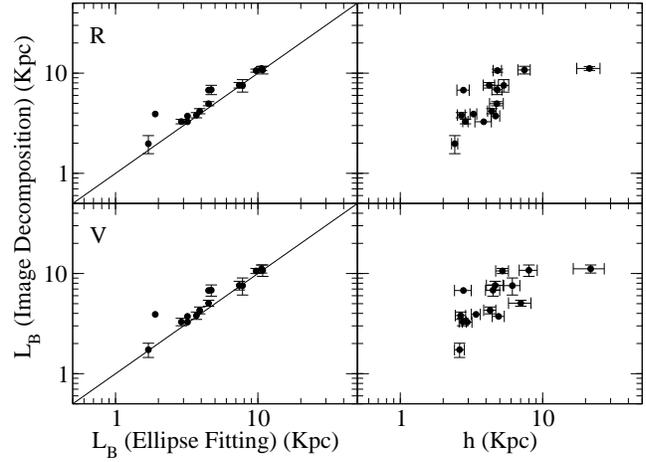}
   \caption{Left: Correlation between the length of bars from image fitting and from the measurements using
ellipse fits in GdS06. The solid lines depict a perfect correspondence. Right: Correlation between the length
of bars from image fitting and the scale-length of discs. The outlying point corresponds to NGC 5850, whose disc
parameters suffer from large uncertainties. These results include all barred galaxies in the
sample, in both bands, as indicated.}
   \label{bars}
\end{figure}

\begin{figure}
   \centering
   \includegraphics[width=7cm,clip=true]{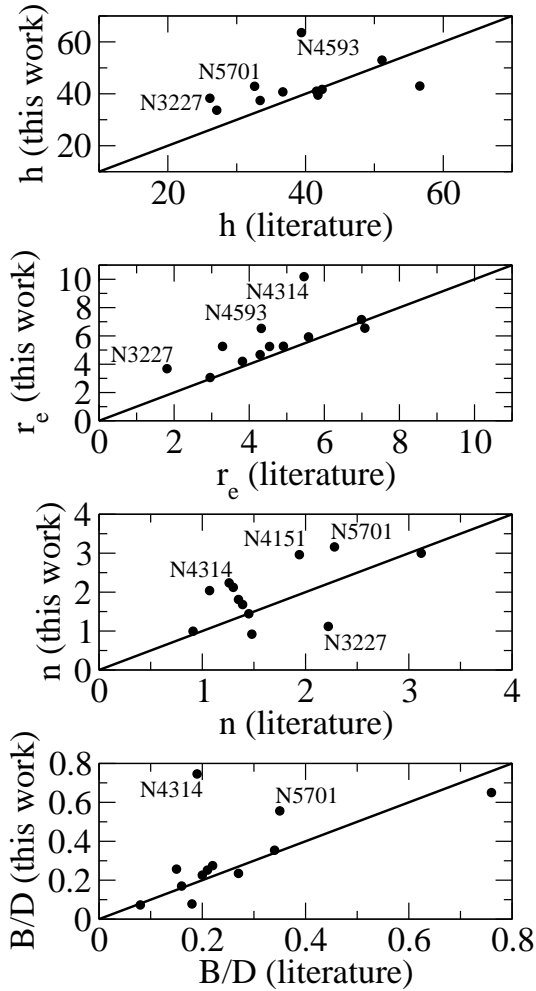}
   \caption{Comparison between some structural parameters obtained in this work with those found in \citet{lau04} and
\citet{lau05} for the same galaxies. From top to bottom: disc scale-length, bulge effective radius, bulge
S\'ersic index and bulge-to-{\em disc} luminosity ratio. Scale-lengths are in arcsec. Some cases where a good agreement
was not found are indicated. The solid lines depict a perfect correspondence.}
   \label{complau}
\end{figure}

\begin{figure}
   \centering
   \includegraphics[width=5.5cm,clip=true]{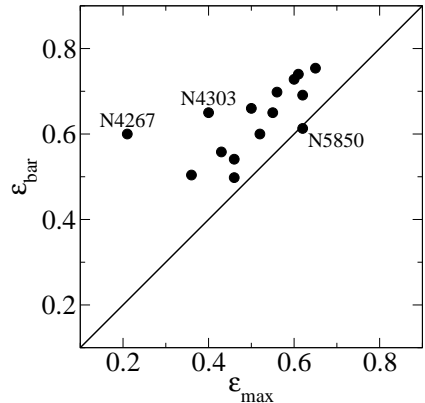}
   \caption{Ellipticity of bars estimated from image decomposition plotted against the ellipticity peak
in ellipse fits from GdS06. Some interesting cases are indicated. The solid line indicates a perfect correspondence.
It is clear that ellipse fits systematically underestimate the true ellipticity of the bar.}
   \label{ecc}
\end{figure}

It is not uncommon to see in residual images such as those in Fig. \ref{results} regions with evident negative
residuals, where the fitted model is brighter than the galaxy. In some cases, this is clearly a result
of dust extinction, but in other cases their presence might mean that the models used are not fully
adequate. Inspecting the results in Fig. \ref{results}, it is possible to identify two particular cases
where such negative residuals are not only conspicuous, but also clearly delineate a distinct region
in the residual image. These are NGC 4608 and NGC 5701. In their residual images, one is able to spot a
region in the disc where the models are definitely brighter than the galaxies. It can be described
as two crescents, one at each side of the bar, but out of it (pointed out by the red arrows in
Fig. \ref{results}). In NGC 4608, these crescents extend to a radius similar
to the bar semi-major axis length, i.e., up to the inner ring surrounding the bar. In NGC 5701, this region
is more extended, and the crescents occupy the whole area between the bar and the outer ring. In
\citet{gad03}, this less luminous area in the disc component of these galaxies was identified, and it was pointed
out that the $N$-body models of bar formation and evolution presented in \citet{ath02} produce a very similar
feature, particularly their models which lead to the formation of very strong bars (see their Fig. 3).
\citet{ath02b,ath03} presents theoretical work which shows that bars get longer with age by capturing disc stars,
which is in agreement with the observational findings in \citet{gad05} and GdS06, although more observational
evidences are needed to settle this point. Thus, these fainter areas in the discs of NGC 4608 and NGC 5701
are plausibly created by such capture of stars. If this is indeed the case, then these galaxies provide
an excellent opportunity to gather insight on these theoretical results and the secular evolutionary
processes in barred galaxies. One of the implications from these studies is that the initially
exponential density profile of a stellar disc in a given galaxy would evolve into a more complex
structure, which can be better described by a density profile which is flatter in the inner part, as compared
to the outer part. Such a profile is reminiscent of a Freeman type II profile \citep{fre70}, which is exponential
in the outer part and flatter in the inner part. The amount of flattening in the inner part of the disc
profile of both NGC 4608 and NGC 5701 can then give us a quantitative measure of these secular evolution processes.

An issue in the results presented in \citet{gad03} was that the model used to fit these galaxies did
not include a bar. In fact, the absence of this component disturbed the fit to the extent that it was
not possible to find a solution for the disc component of both NGC 4608 and NGC 5701.
\citet{lau05} investigated this issue in both galaxies and,
using a model that includes a bar, found similar results as the ones shown in Fig. \ref{results}.
After accounting for the bar, a solution for the disc is possible, but the negative residuals in the inner
disc still remain significant.

\begin{figure}
   \centering
   \includegraphics[width=5.5cm,clip=true]{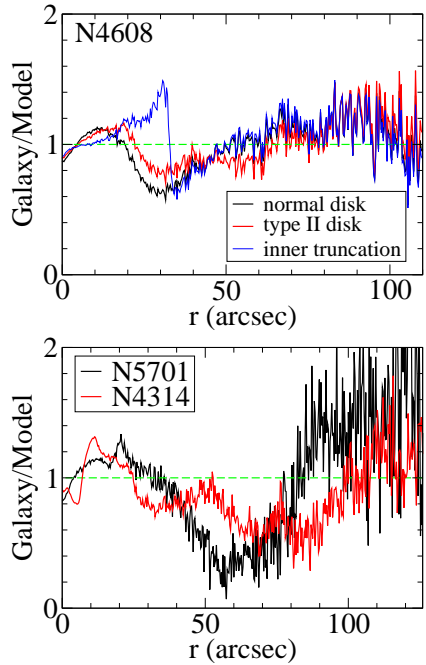}
   \caption{Radial profiles of the residuals from the three different models for NGC 4608 (top panel) and
from the models for NGC 5701 and NGC 4314 (bottom panel).
These profiles were built with the residual images along a stripe
perpendicular to the bar, and with a width of five pixels, from which an average value was taken using both sides
from the centre at each galactocentric radius. One sees that the best fitting model for NGC 4608 is the one where
the disc is a type II disc (see also Fig. \ref{ftru}),
rather than the usual disc, that goes exponentially all the way to the
centre (type I), or the disc with an abrupt inner truncation. In addition, the area of the disc in NGC 5701 between the
bar and the outer ring is about $2-3$ times fainter than the model. As a comparison, on can also see negative residuals
in the disc between the bar and the spiral arms in NGC 4314, but it is a less pronounced effect. In all the other
galaxies in the sample such negative residuals are not conspicuous.}
   \label{resp}
\end{figure}

\begin{figure*}
   \centering
   \includegraphics[height=4cm,clip=true]{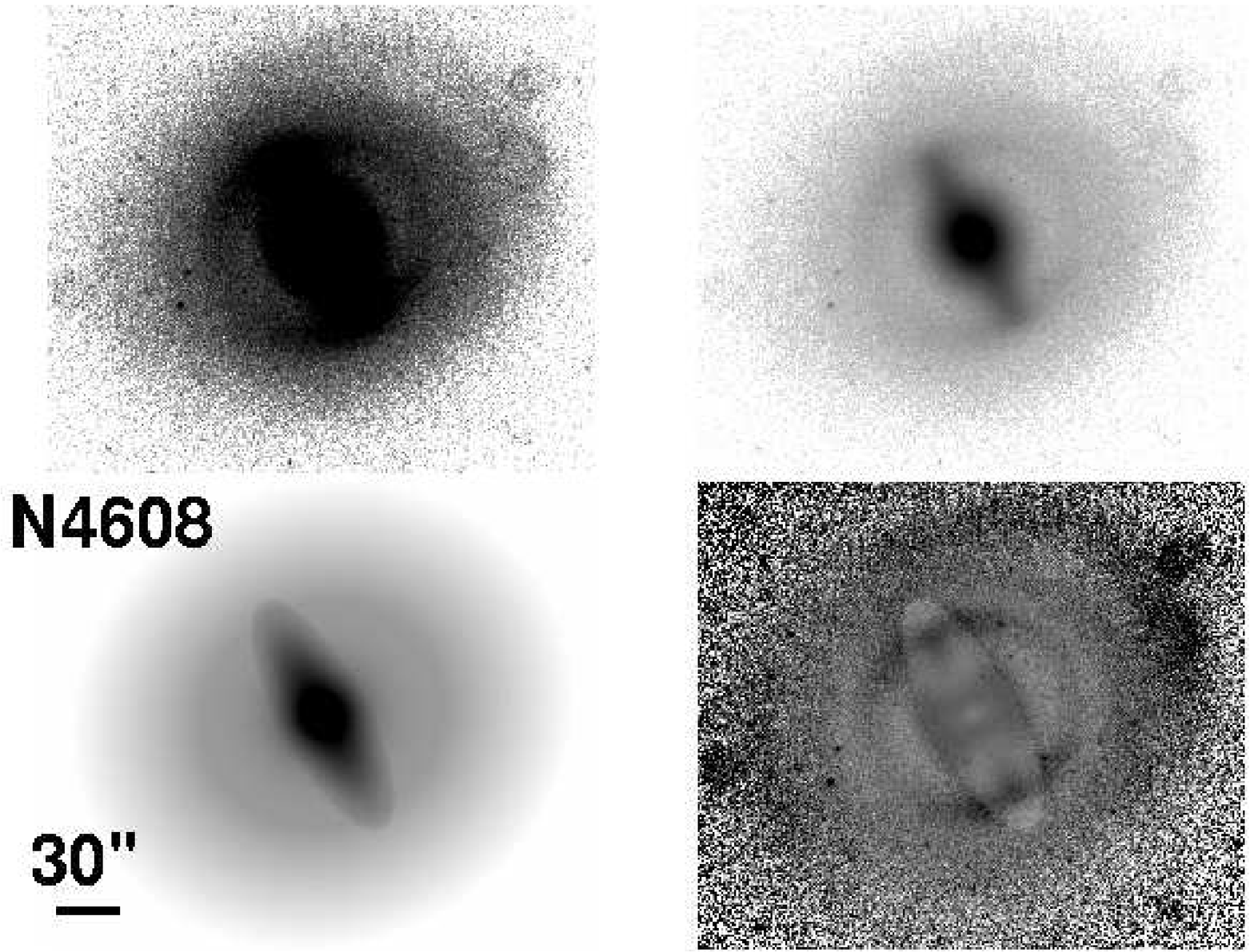}
   \includegraphics[height=4cm,clip=true]{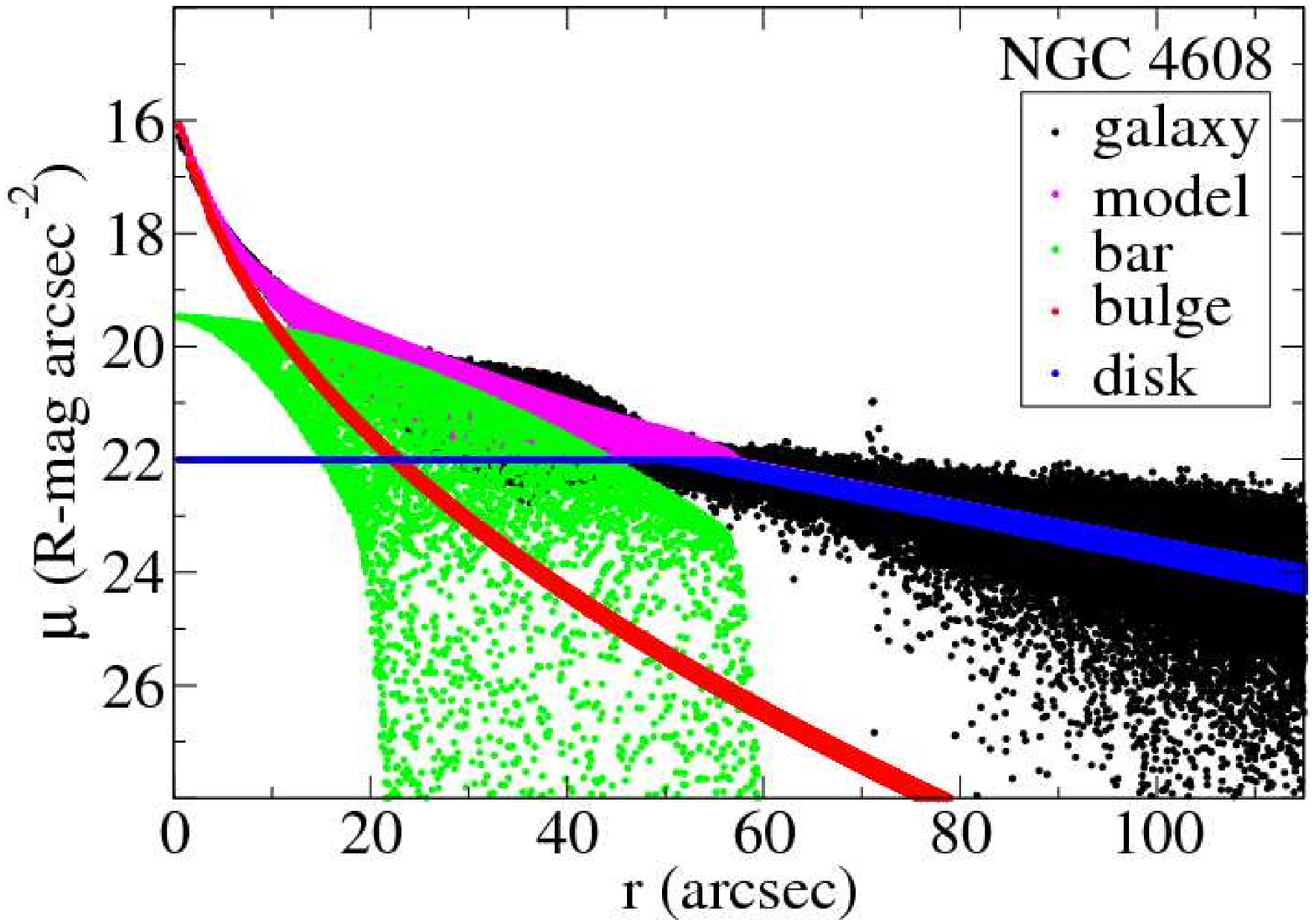}
   \includegraphics[height=4cm,clip=true]{ftruc.eps}
   \includegraphics[height=4cm,clip=true]{ftrud.eps}
   \caption{Similar to Fig. \ref{results}, but when the disc in NGC 4608 is modelled as a Freeman type II disc.
Comparing the results obtained from both models it is clear that a type II disc produces a better fit to this galaxy.}
   \label{ftru}
\end{figure*}

In order to obtain a better fit to NGC 4608, i.e., one in which the match between model and galaxy is
better in the inner, fainter part of the disc, and thus does not produce such strong negative residuals,
and at the same time be able to quantify how faint the inner part of the disc is compared to what would be
expected from a single exponential profile for the whole disc,
I tried another two models for the disc component. In both models, the disc luminosity profile is exponential,
as before, but only from a certain radius outwards. The difference resides in the behaviour of the profile from
this point inwards: in the first model, the profile is abruptly truncated, it falls to zero, and so the disc
has a hole in the centre; in the second model, the profile stops rising exponentially and is kept at a
constant value all the way to the centre, so that the disc is similar to a Freeman type II disc
(note that this model has a much stronger physical justification than the former).
The fitting in these cases was done keeping the bar component as found in the best fit using a type I disc.
The radius at which the disc profile changes is left as a free parameter.
The top panel in Fig. \ref{resp} compares the three models. It shows radial profiles built from the
residual images along a stripe perpendicular to the bar, and with a width of five pixels, from which an average
value was taken using both sides from the centre at each galactocentric radius. Since the residual images
are built dividing the galaxy image by the model image, values above one in these profiles indicate that
the galaxy is brighter than the model at the corresponding pixels, whereas values below one indicate that
the model is brighter. The fainter, inner part of the disc in NGC 4608 can easily be identified in this figure,
as a depression between $r\approx20\arcsec$ and $r\approx50\arcsec$. One sees that, at the minimum of this
depression, the model with a normal, type I disc is almost a factor of two brighter than the actual disc.
On the other hand, using the model with a hole in the disc does not result in a better fit, since in this
case one sees now a significant positive residual between $r\approx20\arcsec$ and the truncation radius.
In the fit with a type II disc, however, the match between model and galaxy is clearly improved. The full results
of this fit are shown in Fig. \ref{ftru} and also in Table \ref{rres}.
The better agreement is evident in the residual image,
in the residuals of the elliptically averaged brightness profiles, and also in the ellipticity profile, where
the steep drop at the end of the bar is now better reproduced. As expected, the value found by the code for the
radius of the break in the disc profile is similar to the bar length (55\arcsec, i.e., $\approx2$\arcsec\ less than
$L_{\rm Bar}$). Note that the brightness profile of the galaxy also has a break at this point. Furthermore,
there is an improvement of $\approx25\%$ in the $\chi^2$ value. Interestingly, the disc luminosity fraction
falls from 0.565 (using a type I disc model) to 0.491 (with a type II disc),
i.e., 7.4\% of the total galaxy luminosity, which is $\approx40\%$ of the {\em bar}
luminosity. Assuming that NGC 4608 had a type I disc initially, and that the stars from the inner disc
migrated to the bar, resulting in the change of the disc profile to a type II, then one can conclude
that the bar increased in mass by a factor of $\approx1.7$, through the capture
of $\approx13\%$ of the disc stars. This assumes the same mass-to-light
ratio in the bar and in the disc, but is reasonable enough for an order of
magnitude estimation. In fact, we showed in GdS06 that the B$-$R colours of the bulge, disc and bar in NGC 4608
are very similar, meaning that the corresponding mass-to-light ratios might not be too different.

For NGC 5701 it was not possible to obtain an alternative fit. When trying both the inner truncated disc
and the type II disc, the radius at which the disc profile changes, as found by the code, is close
to the centre, and thus the resulting model is not significantly different from the previous one.
A possible explanation for this difference is that in NGC 4608 the two crescents together form
a well defined single region with negative residuals, whereas in NGC 5701 the two crescents are more
detached from each other, resulting in two separate regions with negative residuals. Thus, it seems
that a more complex model is necessary. In Figs. \ref{results} and \ref{resp} one sees that the
mismatch between galaxy and model in the inner disc of this galaxy is even more accentuated than in the
case of NGC 4608. The area of the disc in NGC 5701 between the bar and the outer ring is about $2-3$ times
fainter than the model. In almost all the other galaxies in the sample such negative residuals are not
conspicuous, the only exception being NGC 4314. Figures \ref{results} and \ref{resp} show, however,
that this effect is less pronounced in this galaxy.

\subsection{The Effects of Neglecting Bars}
\label{sec:nobar}

In order to study the effects of not modelling bars on the structural parameters obtained from image decomposition,
the fitting of the barred galaxies in the sample was repeated with the bar removed from the models. As I will shortly
show, bulge models are significantly altered when bars are not taken into account, to accommodate the light
from the bar. Thus, in this exercise,
I fixed the ellipticity and position angle of the bulge, with the results found in Sect.
\ref{sec:decomps}, {\em minimising} the distortion in the bulge models. This means that the effects
caused by ignoring bars, as found here, are actually lower limits. Apart from these differences,
the fitting process was identical to that of the main fits. This exercise was done
only with the R-band images, and excludes the five galaxies in the sample
for which the AGN contribution has to be modelled, to avoid complicating the interpretation of the results.

\begin{figure}
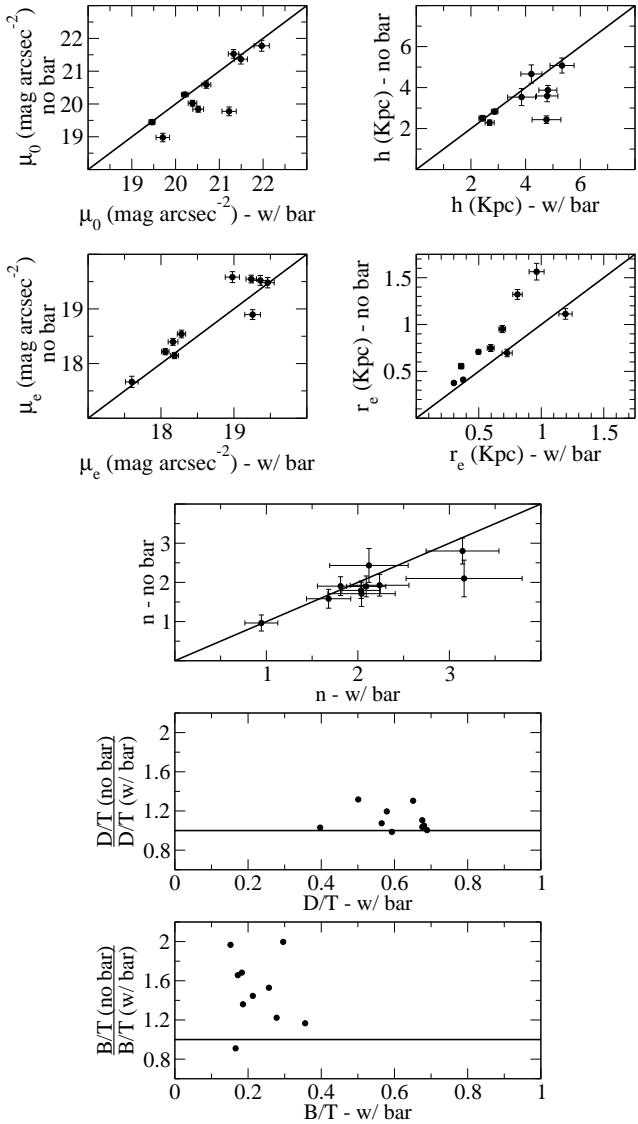

   \centering
   \includegraphics[width=8.4cm,clip=true]{nobara2.eps}\\
   \vskip 0.25cm
   \includegraphics[width=6cm,clip=true]{nobarb2.eps}
   \caption{Structural parameters of discs and bulges,
as estimated when bars are not included in the models, plotted against the same parameters when bars are
taken into account. The two bottom panels show the relative overestimation of the disc-to-total and bulge-to-total
luminosity fractions when bars are neglected, plotted against the corresponding parameters when the models
include bars. The solid lines depict a perfect correspondence. This analysis excludes the five galaxies in the
sample for which the AGN contribution has to be modelled, to avoid complicating the interpretation of the results.
It shows that when bars are ignored, discs tend to assume steeper luminosity profiles, and bulges get bigger, in a
way to accommodate the light from the bar. The effect is stronger for bulges: the bulge luminosity fraction can be
overestimated by a factor of two.}
   \label{nobar}
\end{figure}

Figure \ref{nobar} compares the structural parameters of discs and bulges, and the disc-to-total and bulge-to-total
luminosity ratios, as estimated when bars are not included in the models, with the same parameters when bars
are taken into account. It is clear that both bulge and disc components are altered in order to accommodate the
light from the bar. Discs tend to assume a steeper luminosity profile, meaning brighter $\mu_0$ and shorter $h$.
As a consequence, the disc luminosity fraction increases. A stronger effect is seen in the bulges, which get
bigger to account for the bar, acquiring larger $r_e$ and luminosity fractions. The changes in $\mu_e$ and $n$
are within the uncertainties but it seems to be a systematic effect towards fainter $\mu_e$ and smaller $n$.
It is interesting to note that in some cases there was no significant change. Evidently, if the bar contributes
to a large fraction of the total galaxy luminosity these effects will be more pronounced. Nonetheless,
other features in the galaxy might have a relevant role as well. For instance, if the geometrical parameters
of the bar are very different from those of both bulge and disc, this will give further constraints to help
the code in order to separate the different components, even if the bar is not modelled. To evaluate the exact
circumstances that aggravate this issue is beyond the scope of this study. The relevant points
to stress here are the presence of {\em systematic} effects at play when bars are not considered in the models, and
the fact that these effects alter primarily the structural parameters obtained for bulges. Figure \ref{nobar}
shows that the disc luminosity fraction is overestimated, on average, by 10\%, with a maximum overestimation
of 30\%. The bulge luminosity fraction is overestimated, on average, by 50\%, and this overestimation can be as high
as a factor of two.

\begin{figure}
   \centering
   \includegraphics[width=6cm,clip=true]{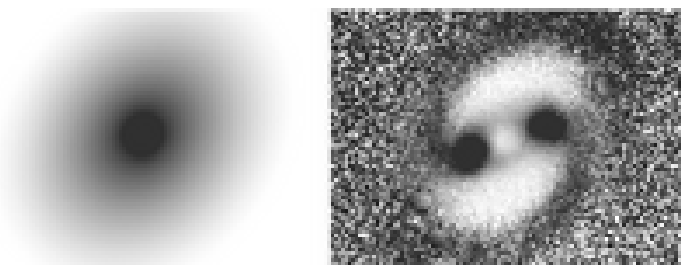}\\
   \vskip 0.25cm
   \includegraphics[width=6cm,clip=true]{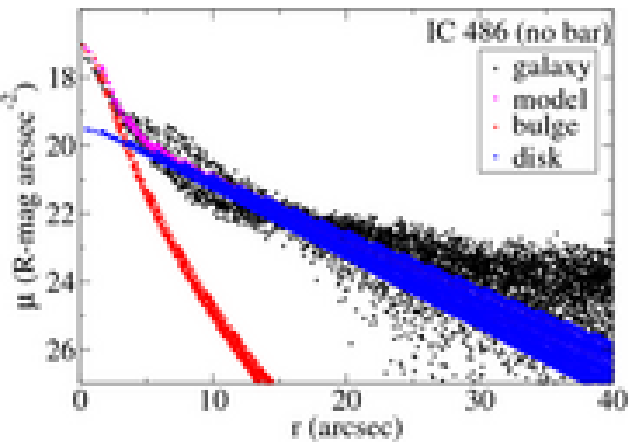}
   \caption{Results from the R-band image decomposition of IC 486 when the bar is not included in the model. Top left:
total model image; top right: residual image; bottom: surface brightness profiles of the galaxy and the models.
The images and the panel shown are similar to the corresponding ones in Fig. \ref{results}, and these should be compared
in order to assess the effects of neglecting the bar. When the bar is not taken into account in the fitted model, the
disc acquires a steeper profile, with a brighter central surface brightness, while the effective radius of the bulge
grows. The resulting bulge-to-total and bulge-to-disc ratios get higher by 45\% and 21\%, respectively. These changes
are reflected in the residual image: the bulge model absorbs the inner part of the bar, and the brightened bulge and
disc models produce strong negative residuals.}
   \label{I486}
\end{figure}

Figure \ref{I486} illustrates the effects of not including the bar in the fitted model in a more detailed fashion,
for an individual case, that of IC 486. This figure should be compared with Fig. \ref{results}, which shows the
results from a fit that includes a bar component in the model. One sees that the disc acquires a steeper profile
($h$ gets shorter by about 25\%), with a brighter central surface brightness (0.73 magnitudes brighter),
while the effective radius of the bulge grows about 25\%. The disc luminosity fraction increases 20\%,
whereas the bulge luminosity fraction increases 45\%. These changes are reflected in the residual image: the bulge
model absorbs the inner part of the bar, and the brightened bulge and disc models produce strong negative residuals.

\subsection{The Effects of Neglecting Bright AGN}
\label{sec:noagn}

For five galaxies in the sample it was deemed necessary to include the AGN component in the model in order
to obtain proper fits. To study the effects of not modelling the AGN on the structural parameters obtained from
image decomposition, the fitting of these galaxies was repeated with the AGN component removed from the
models. Again, apart from that, the fitting process was identical to that of the main fits, and only R-band
images were used.

\begin{figure}
   \centering
   \includegraphics[width=8.4cm,clip=true]{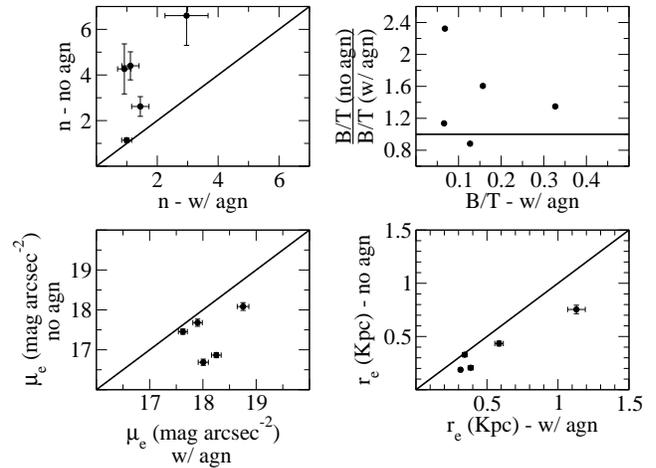}
   \caption{Structural parameters of bulges, and the relative overestimation of the bulge-to-total luminosity ratio,
as estimated when AGN are not accounted for in the models, plotted
against the corresponding parameters when AGN are taken into account. The solid lines
depict a perfect correspondence. It is clear that, due to the concentrated light from the AGN, the bulge parameters
can not be reliably retrieved for galaxies hosting bright AGN if its contribution is not modelled. In this case, bulges
tend to become smaller and more luminous. The S\'ersic index of the bulge is the most affected parameter and can be
overestimated by a factor of four. The bulge luminosity fraction can be overestimated by a factor of two.}
   \label{noagn}
\end{figure}

Figure \ref{noagn} compares the structural parameters of bulges, and the bulge-to-total luminosity ratio, as
estimated when AGN are not accounted for in the models with the same parameters when AGN are taken into
account. The disc and bar components are not significantly affected when ignoring the AGN contribution.
The figure shows that, due to the concentrated light from the AGN, if it is not taken into account, bulges
tend to become smaller and more luminous, i.e., with shorter effective radius and brighter effective
surface brightness. Most importantly, the S\'ersic index of the bulge is severely affected, being overestimated
by up to a factor of four. As a result of these changes, the bulge luminosity fraction is also overestimated,
up to factor of two. For two of these galaxies, NGC 4303 and NGC 4579, where the AGN component corresponds to only
$\approx1\%$ of the total galaxy light, these effects are small. For the remaining three, NGC 3227, NGC 4151
and NGC 4593, where the AGN luminosity fraction ranges from $\approx4\%$ to $\approx9\%$, these effects are
significant. In addition, although the number of points is small, these changes seem to be systematic.
Thus, it is clear that the bulge parameters can not be reliably retrieved for galaxies hosting bright AGN if its
contribution is not modelled.

In this study, the decision on whether or not to include an AGN component in the model was essentially based on the
galaxy surface brightness profile and the AGN classification. A cuspy profile in a galaxy hosting an AGN indicates
that the AGN contribution has to be taken into account. For larger studies, it would be interesting to have
a way to predict if a galaxy needs the AGN component in the fit, without having to inspect the galaxy surface
brightness profile. In principle, one would include the AGN model if the galaxy is a type 1 AGN. For one of the
type 1 AGN galaxies in the sample, however, the AGN component was not needed. This might not be a matter of
too much concern, though, as the code brought down the AGN contribution to the total galaxy light to negligible
values, when the first fit tried included this component. The resulting fit was not substantially different
from the final fit, with the AGN removed from the model. On the other hand, weaker AGN, like in NGC 4579, might
need an AGN component for an accurate fit, but, as shown above, the effects of neglecting the AGN in the
fit to this galaxy are small. It thus seems that the fitting procedure is quite robust in deciding how important
is the AGN contribution to the galaxy light distribution.
This likely results from the fact that the AGN model contains only one free parameter, its peak intensity,
plus one fixed parameter, the FWHM of its profile, which is given by the seeing FWHM.
This suggests that it might not be too harmful,
at least statistically, if an AGN component is included in the fits of all galaxies classified as AGN,
provided that the input values for the bulge parameters given at the initial setup of the fit are
reasonable. A more appropriate procedure could be to evaluate the effects of dust obscuration in the centre
of the galaxy, since heavily obscured AGN probably do not need to be modelled. With this aim, parameters
such as the column density of absorbing Hydrogen and the ratio between far-infrared and optical or ultra-violet
luminosities could be useful.

\subsection{The Light from Spiral Arms}
\label{arms}

By inspecting the original and residual images in Fig. \ref{results}, it is possible to see that many of the
galaxies in the sample have bright spiral arms, which, in some cases in fact, appear to be a dominant component.
Should these be taken into account in the image decompositions? To answer this
question, one must first keep in mind that, to produce Fig. \ref{results}, the original images were displayed in
a logarithmic scale, thus emphasising structures with low surface brightness. Furthermore, the residual images
were displayed with a very narrow dynamic range, also in order to point out faint, residual sub-structures.
Thus, at least from a qualitatively viewpont, the impression that the spiral arms might be a dominant component
in these galaxies can be possibly wrong (at least in some cases), and results from the approach often used to render
galaxy images and residual images.

A possible way to quantify what is the fraction of the galaxy light that comes from the spiral arms involves
using the residual images. To this end, I produced new residual images, which were made by subtracting the model
image from the original image (rather than dividing the latter by the former, as done before). The mean pixel
value and the corresponding standard deviation were calculated within a region comprising the whole galaxy, i.e.,
from the centre to where the spiral arms end, as seen in the residual image. The mean value is always close
to zero, which reveals an important aspect of the fitting procedure and the residual images: when fitting the disc
component to a galaxy with spiral arms, the code tries to minimise the deviation between the galaxy and the model,
thus creating an average disc model, which, on the one hand, accounts for at least part of the light that
comes from the arms, and, on the other hand, produces slightly negative residuals in the inter-arm region.
These negative residuals can be seen in Fig. \ref{results}, and should not be confused with the much more
negative residuals discussed in Sect. \ref{4608}. The important point is thus that at least part of the light from
the spiral arms is already accounted for in the disc component of the model. With the residual images one can thus
quantify how much light is in excess and was not included in the model. This was done by dividing the mean pixel
value within the galaxy region in the residual image by the corresponding value in the original image, after
statistically eliminating pixels with exceedingly negative values (i.e., more than three times the standard
deviation -- it turned out that this procedure does not significantly change the results). This was done
for all galaxies, except NGC 2110, NGC 2911 and NGC 4267, where no conspicuous spiral arms are seen. No
systematic or significant
difference is seen when comparing the results from the R-band and the V-band images, but since the
former have higher signal-to-noise ratio I consider the results from these images more reliable
to discuss this issue. For IC 486,
NGC 4477, NGC 4593 and NGC 5850, it was not possible to obtain a reliable value, i.e., the excess light
is consistent with being zero, or at least only a tiny fraction of the total galaxy light
($<0.1\%$). For most
galaxies the excess light is $\approx1-2\%$ of the total galaxy light. For three galaxies the excess light
is significant: NGC 4303 (10\%), NGC 4314 (4\%) and NGC 5383 (4\%).

Thus, these results suggest that, for most disc galaxies, the inclusion of a spiral arm component in the
model is not a fundamental issue, and that the fraction of stellar mass contained within the spiral arms
is too small, especially considering that the average mass-to-light ratio of the young stars in the arms is
likely to be substantially lower than that in the remaining of the disc and the galaxy. Nevertheless, for
some disc galaxies, particularly those with late morphological types, this fraction might be relatively
important. A fully satisfactory approach to tackle this issue would thus be to include a spiral arm
component in the model, resorting to, e.g., Fourier techniques. One would then be able to answer
more adequately what is the fraction of the disc and galaxy stellar mass that resides in spiral arms.
This is, however, beyond the scope of this study.

\section{Decomposition of Redshifted Images}
\label{sec:hz}

Recently, research on the structural properties of galaxies has shifted the main focus from the very nearby
universe to samples of more distant galaxies, as described in the Introduction, provided by surveys such as the
SDSS \citep[see, e.g., ongoing work in][]{gadlp}. This makes possible to carry on statistically significant analyses
and, in some such studies, evaluate how the structure of galaxies changes with time. In many of these studies,
however, the physical spatial resolution achieved in the galaxy images is substantially lower than what
can be routinely achieved for nearby galaxies. It is thus most relevant to ask what impact such diminished
resolution can have on the results. In addition, given the results from the previous section, one can ask
whether bar and AGN components are still needed in the models when pushing to such low resolution regimes.
These questions are addressed in this section.

\subsection{Fitting Low Resolution Images}
\label{sec:hzgen}

\begin{figure*}
   \centering
   \includegraphics[width=17cm,clip=true]{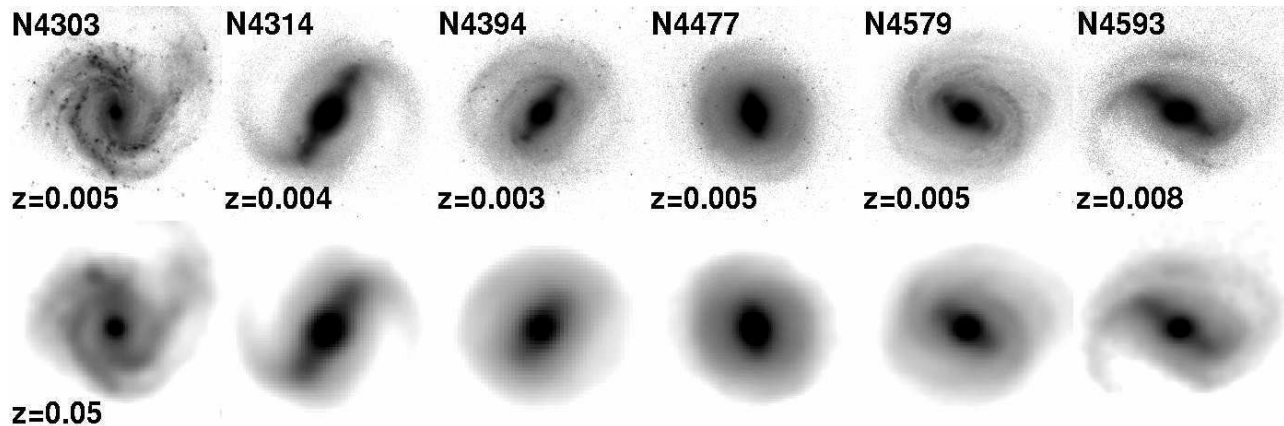}
   \caption{Comparison between the original images (top row) and artificially redshifted images (bottom row) for
six galaxies in the sample, as indicated. The redshifted images simulate how the galaxies would look like if located
at a redshift $z=0.05$, i.e., $\approx10$ times farther than their actual location, and observed with a seeing of
1.5\arcsec.}
   \label{hzimgs}
\end{figure*}

To examine the effects of using low resolution images for galaxy image decomposition, the original
images used above were artificially redshifted to $z=0.05$, which is $\approx10$ times farther than the actual
location of the galaxies. This was done by demagnifying the images by the appropriate factor using the
task {\sc magnify} in {\sc iraf}, keeping the pixel angular size.
In order to have the same resolution in all redshifted images, these were
also convolved with a circular Gaussian, using the task {\sc gauss},
in such a way that the final resolution is 1.5\arcsec,
taking into account the resolution in the original image. This is the median seeing FWHM in the SDSS and, at
$z=0.05$, corresponds to a physical spatial resolution of $\approx$ 1.5 Kpc. Such a resolution is typical
in other works as well \citep[e.g.,][]{all06}, and is what can be achieved at $z\sim1-2$ in images from
HST, or from ground-based telescopes with adaptive optics.
Thus, the following results have a broad applicability, being relevant to studies of the structure of galaxies
in both the local universe, like the ones just mentioned, and at higher redshifts, such as COSMOS
\citep[e.g.,][]{koe07} and SINS \citep[e.g.,][]{gen06}.
A comparison between the original and redshifted images proves very instructive. This is done in Fig. \ref{hzimgs}.
Features such as the star forming knots and dust lanes in NGC 4303 are completely smoothed out, and only
a hint of the spiral arms in NGC 4394 and the bar in NGC 4477 can be seen at low resolution.

\begin{figure}
   \centering
   \includegraphics[width=8.4cm,clip=true]{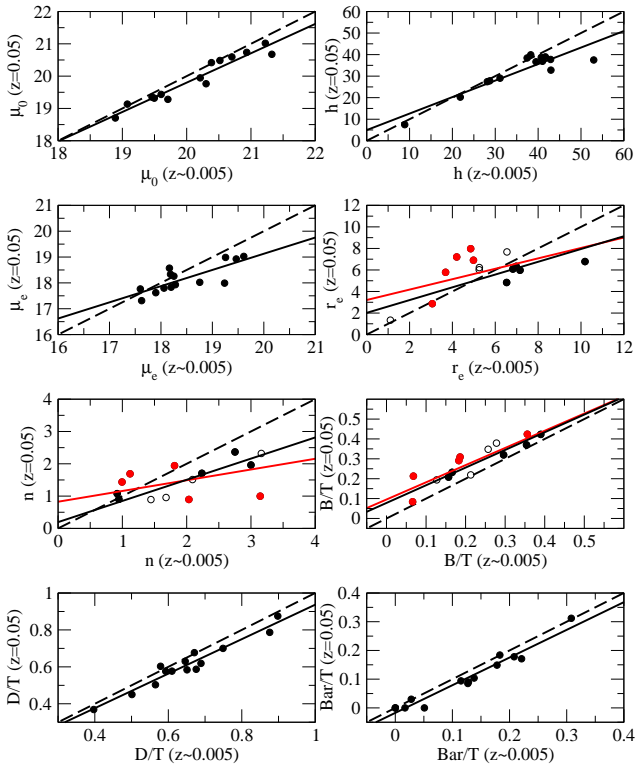}
   \caption{Structural parameters of discs and bulges, and bulge, disc and bar luminosity fractions, as determined
with the redshifted images, plotted against the same parameters obtained with the original images. The dashed lines
indicate a perfect correspondence. Luminosity parameters are in units of mag arcsec$^{-2}$ and scale-lengths in
arcseconds. For a proper comparison with the original images, the scale-lengths from the redshifted images are
scaled back to the original galaxy distance. For $r_e$, $n$ and B/T, filled circles correspond to those galaxies
where the effective radius of the bulge in the redshifted image is larger than the seeing radius, the empty circles
correspond to those galaxies where it is similar to the seeing, and the red points correspond to those galaxies
where it is smaller than the seeing. The solid lines are linear fits to the data. For $r_e$, $n$ and B/T, the red
lines are fits to all data points, while the black lines correspond to fits where the red points were excluded.
The parameters obtained from these linear regressions are shown in Table \ref{regress}.
One sees that, in general, structural parameters can be reliably retrieved through
image fitting even in the low resolution regime. Nevertheless, bulge parameters are prone to errors if its
effective radius is small compared to the seeing radius, and might suffer from systematic effects.}
   \label{hz}
\end{figure}

The same procedures applied during the fitting of the original images were repeated with the redshifted images.
Due to the larger uncertainties in the images of NGC 4151, NGC 4665 and NGC 5850,
these galaxies were excluded from this analysis.
It should be noted that, to allow a fair comparison between the results from both sets of images, the
results obtained with the original images were {\em not} used to constrain the fitting of the redshifted
images. In Fig. \ref{hz}, the structural parameters of discs and bulges, and the bulge, disc and bar luminosity
fractions, as determined with the redshifted images, are compared with the same parameters as obtained with the
original images. For a proper comparison with the original images, the scale-lengths from the redshifted images
are scaled back to the original galaxy distance. One sees a very good agreement in what concerns the disc
parameters and the disc and bar luminosity fractions. The agreement is also quite reasonable
for the effective surface brightness of
the bulge, but less so for its effective radius, S\'ersic index and luminosity fraction.
To understand better why some bulge parameters were not satisfactorily retrieved, the bulges were separated
according to their effective radius in the redshifted images. Thus, in the corresponding panels in Fig.
\ref{hz}, filled circles refer to those galaxies where the ratio of the effective radius of the bulge in
the redshifted image to the seeing radius (0.75\arcsec) is between $\approx1$ and
$\approx2$; the empty circles correspond to those galaxies where this ratio is between $\approx0.8$ and
$\approx0.9$, and the red points correspond to those galaxies where it is between $\approx0.4$ and
$\approx0.6$. Clearly, the worst discrepancies almost always occur when the effective radius of the bulge
is considerably small compared to the seeing radius. When the former is similar or larger than the latter the
agreement between the results from original and redshifted images is somewhat improved.

Figure \ref{hz} also shows lines that are linear regressions to the data points.
The parameters that describe these lines, and their statistical uncertainties, and the corresponding correlation
coefficients are shown in Table \ref{regress}. Although the sample is relatively small, from these fits
it is possible in principle
to evaluate if there are systematic effects in the results in the low resolution regime, and
which structural parameters are most robust. Within the uncertainties, one sees that there seems to be no systematic
effects in the determination of $\mu_0$, $h$, D/T and Bar/T, and thus these parameters can be determined
very reliably, in particular the central surface brightness of the disc, $\mu_0$. As expected, bulge parameters
are the most affected, even after removing those with $r_e$ small compared to the seeing radius. There seems to be a
systematic effect in $\mu_e$, in the sense that bulges fainter than $\approx$ 18.5 R-mag arcsec$^{-2}$
are retrieved with a somewhat brighter $\mu_e$ when using the redshifted images. Similar effects seem to happen
with $r_e$ and $n$: the redshifted images provide smaller bulges if $r_e$ is larger than about 6 arcsec, and
less centrally concentrated bulges if $n$ is greater than around 2. Note, however, that the typical 1 $\sigma$ error
given by {\sc budda} for $n$ is $\approx0.5$, which is quite big considering the full range covered by this
parameter (e.g., here, only from $\approx1$ to $\approx3$).
This complicates the use of the bulge S\'ersic index for a quantitative  morphological
classification of galaxies. To this aim, the bulge-to-total ratio seems to be a more robust parameter, since
it shows the highest correlation coefficient amongst the bulge parameters.
It also shows, however, a somewhat clearer systematic
effect: bulge-to-total ratios recovered in the low resolution regime are systematically larger than those estimated
with the original images. Nevertheless, it is possible to suggest a mean correction from the data in Table
\ref{regress}. The corrected B/T, as a function of the estimated B/T (estimated in a low resolution regime), is given
by:

\begin{equation}
{\rm B/T}_{\rm corr}\approx1.124\times{\rm B/T}_{\rm est}-0.090.
\end{equation}

\noindent From the data in Fig. \ref{hz}, one sees that the average overestimation of B/T, due only to the low
physical spatial resolution, is $\approx5\%$ of the galaxy total luminosity. This light fraction, of course,
has to be redistributed to the other galactic components. Although, as mentioned above, D/T and Bar/T
do not show statistically significant systematic effects, one sees that most points in the corresponding panels
in Fig. \ref{hz} lie close to, but below
the perfect correspondence line, which makes this picture globally consistent.
It is worth stressing again, though, that these assessments are based on a small sample and must be used
with this caveat in mind.

\subsection{The Effects of Neglecting Bars and Bright AGN at Low Resolution}
\label{sec:hzbaragn}

\begin{figure}
   \centering
   \includegraphics[width=8.4cm,clip=true]{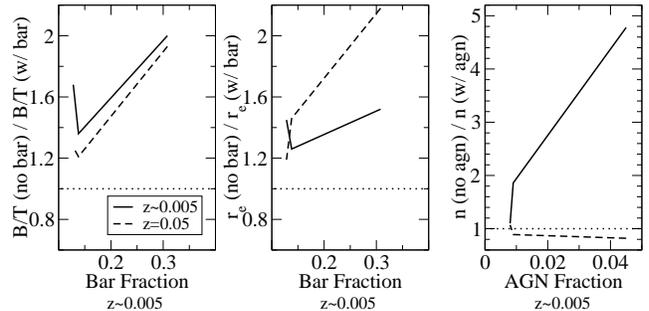}
   \caption{Left and central panels: Overestimation of the bulge-to-total luminosity ratio and the effective
radius of the bulge, as a function of the bar luminosity fraction, when there is no bar in the fitted model,
relative to the same parameter when the bar is taken into account.
Right: Overestimation of the bulge S\'ersic index as a function of the AGN luminosity fraction,
when the AGN light is not modelled, relative to the same parameter when the models include AGN,
for galaxies where an AGN component is included in the fit of the original images (see Section \ref{sec:hz}).
The solid lines refer to the original images while the dashed lines refer to the artificially redshifted
images. The dotted lines indicate no change in the parameters.
This figure is a graphical representation of Tables \ref{nobart} and \ref{noagnt}. It shows that the
overestimation of B/T when ignoring bars, as seen in Sect. \ref{sec:nobar}, is still significant even in the
low resolution regime, but has its strength reduced in this regime if the bar is weak. The corresponding overestimation
of $r_e$ is even considerably worse in the low resolution regime if the bar is not too weak. Furthermore,
it also shows that the overestimation of $n$, when the AGN contribution is not taken into account,
as seen for the original images in Sect. \ref{sec:noagn}, is completely absent in the low resolution regime.}
   \label{hzgraph}
\end{figure}

Using the original images, we have seen that if one does not include bars and AGN in the models,
when fitting galaxies that clearly host such components, the determination of the structural
parameters can be severely affected. However, it is not clear if this is still true when the resolution
of the images used is relatively poor. Given that finer details are smoothed out in this case, one expects
these effects to be less significant, but do they completely disappear? To verify that, a similar exercise
as done with the original images was repeated with the redshifted images. Thus, I selected three of the ten
barred galaxies in the sample with no conspicuous AGN component, which have a reliable original image and
corresponding fit, and span a relatively wide range in bar luminosity fraction,
and fitted a model to their redshitfed images containing
only bulge and disc. As in Sect. \ref{sec:nobar}, the ellipticity and position angle of the bulge were kept
fixed at the values found in the first fit, with the complete model, to the redshifted image,
{\em minimising} the effects caused by the absence of
a model for the bar. Since the major effects of neglecting the bar are on the effective radius of the bulge and
the bulge luminosity fraction, the analysis is focused on these parameters. Table \ref{nobart} compares,
for each of these galaxies, the effective radius of the bulge and the bulge luminosity fraction, as determined
with the original images and the redshifted images, with and without a bar in the model. One sees that,
in cases that bars are prominent, like in NGC 4314 and NGC 4394, the overestimation of these parameters
when the bars are neglected occurs using the redshifted images as much as when the original images were used
(see also Fig. \ref{hzgraph}).
The relative change in the B/T value after omitting the bar is similar in both the original and redshifted
images. The change in $r_e$ is even more pronounced in the latter (by a factor of $\approx2$),
and this might be due to the fact that the geometrical properties of the bulge are in this case substantially
affected by the PSF, which tends to make the bulge rounder, making it harder to
contrain its properties. For a galaxy with a less prominent bar, like NGC 4477, these effects are still
present, albeit with a reduced strength.

Similarly, three of the five galaxies in the sample which were originally fitted with an AGN component in the
model were selected, and their redshifted images fitted without the AGN. Since the major effect of neglecting
the AGN is on the S\'ersic index of the bulge, the analysis is focused on this parameter. Table \ref{noagnt}
compares, for each of these galaxies, the S\'ersic index of the bulge, as determined with the original images
and the redshifted images, with and without the AGN in the model. One sees that the overestimation of the
S\'ersic index, that happens when not accounting for the AGN contribution when fitting the original images,
does {\em not} occur in the low resolution regime, not even in the case of NGC 4593, which has a very luminous
AGN (see Fig. \ref{hzgraph}).

\section{Implications for Studies on the Structural Properties of Galaxies}

Image fitting of galaxies is a complex endeavour, especially when dealing with galaxies rich in structure,
such as barred galaxies. Thus, the fact that with the structural parameters obtained here one can reproduce
previously known correlations
(Figs. \ref{korm}, \ref{disks} and \ref {bars}), and the agreement between these parameters and those from
similar studies in the literature (Sect. \ref{sec:complau}, Fig. \ref{complau}) are very encouraging.
Likewise, the fact that the structural parameters obtained with the redshifted images agree with those obtained
with the original images, as seen in the previous section, gives support to studies
based on more distant galaxies, if the effective radius of the bulge is not small compared to
the PSF, and the physical spatial resolution is 1.5 Kpc or better. With the work presented here one can not
conclude on whether a similar agreement emerges if the resolution is poorer. A word of caution should be
given, however: the redshifted images were fitted and checked individually, and automated procedures usually
applied to large samples normally lead to larger uncertainties. Furthermore, only the effects of a lower
spatial resolution in the images of more distant galaxies are mimicked in the redshifted images, but
other issues, such as dimming and wavelength shifting, might as well be relevant, especially if reaching
$z\sim1$.

The bar luminosity fraction of the galaxies in the sample range,
in the R-band, from around 2 to 30 per cent, with
a median value of $\approx14\%$ and standard deviation of $\approx8\%$. Similar results are obtained from the
V-band images. This broadly agrees with the findings of \citet{gadlp} with a sample of about 100 barred
galaxies, namely, $0.01\leq{\rm Bar/T}\leq0.3$, with a median value $\approx10\%$ \citep[see also][]{ree07}.
This means that the effects of not modelling the bars in barred galaxies, seen in Sect. \ref{sec:nobar},
should be typical. We saw that the most affected parameters are the effective radius of the bulge and the
bulge-to-total luminosity ratio, both being significantly overestimated. We also saw that these effects
hold in the low resolution regime (Sect. \ref{sec:hzbaragn}, Table \ref{nobart} and Fig. \ref{hzgraph}),
provided that the bar is prominent enough. For NGC 4477, the galaxy
with the least prominent bar, amongst the galaxies with which this analysis was done, with ${\rm Bar/T}=0.128$,
these effects are substantially less pronounced in the redshifted images, compared to the original images.
Thus, even at low resolution, these effects are important for at least about half of the barred galaxies,
i.e., roughly about 1/3 of disc galaxies. On the other hand,
it seems reasonable to conclude that, for galaxies with Bar/T below $\approx0.1$, and in the low
resolution regime, the effects of neglecting the bar are within the uncertainties. Nevertheless, even
for these galaxies, such effects should result in a {\em systematic} overestimation of $r_e$ and B/T.

Evidently, regardless of image resolution, ignoring bars in barred galaxies affects results on the stellar
mass budget in the universe, i.e., the distribution of mass in
stars in the different galactic components. When bars are not somehow taken into account, the amount of mass
in stars in bulges and discs is overestimated, and the excess is an indication of the amount of mass in stars
that reside in bars. Using the original images, we have seen that the average overestimation of the bulge and
disc luminosity fractions is, respectively, 50\% and 10\%. Applying image decomposition techniques to
a sample of more than $10^4$ galaxies, \citet{dri07} estimated that the $z\approx0$
stellar mass content in classical bulges and discs is $26\pm4\%$ and $58\pm6\%$, respectively
\citep[see also][]{tas05}. They also find that the stellar mass content in elliptical galaxies is $13\pm4\%$.
Bars are not taken into account in the fitted models, which contain only bulge
and disc. These authors made a thorough quality control, removing a significant fraction of the fits,
which were deemed poor. Thus, one could argue that barred galaxies have not passed quality control.
However, it is usually the case that, even when there is no bar in the model,
when fitting a barred galaxy, one gets an acceptable (though wrong) fit, essentially because the
bulge model is distorted, trying to accommodate the bar light, as shown in Sect. \ref{sec:nobar}.
Hence, we can use their results to obtain a {\em rough} estimate of what can be the stellar
content in bars, assuming that the biases produced by ignoring bars, as found in Sect. \ref{sec:nobar},
can be used in this case to obtain the true bulge and disc luminosity fractions. If one uses the
average result from Eq. (5),
to correct the bulge fraction due to the effects of low spatial resolution, as discussed above (the physical
spatial resolution of the images used by Driver et al. is on average very similar to that of the redshifted
images I use here), and assumes that the fraction of barred galaxies, considering only disc galaxies,
is $\approx70\%$, then the stellar content in classical bulges and discs
is found to be $\approx13.5\%$ and $\approx58.5\%$, respectively. And the stellar content in bars is
$\approx12\%$\footnote{To get to these numbers, first one has to remove 5\% from the original
stellar content in bulges, given by Driver et al., since this is the average overestimation of B/T due
only to the low physical spatial resolution in the images, and add that 5\% to the original
content in discs (one could distribute this fraction between discs and bars, but let's be conservative).
Then, multiply the assumed fraction of barred galaxies, 70\%, by the average overestimation of
B/T due to the absence of bars in the models, 50\%, and multiply the result by the bulge content just found, 21\%.
This results in $\approx7.5\%$, which also has to be removed from this bulge content and added to the {\em bar}
content, which was, up to now, zero. Similarly, for the discs: $70\%\times10\%\times63\%\approx4.5\%$, which
has to be removed from the disc content and added to the bar content.
If one deems the first step unnecessary, the stellar mass content in bulges, discs and bars change to
17\%, 54\% and 13\%, respectively.}.
Incidentally, this fraction of the stellar mass in discs is still very similar to that given by Driver et al.
For bulges, however, the corresponding difference is $\approx3\sigma$, considering their error estimate.
Bulges, discs and bars can have different mass-to-light ratios and this is {\em not} taken into account here,
adding more uncertainty to these estimates. Nevertheless, they open up the possibility of
{\em bars being as important as classical bulges and ellipticals in what concerns their stellar mass content
in the local universe}.

As discussed in Sect. \ref{arms}, the stellar mass content in spiral arms seems to be small, at least for
early-type disc galaxies, and might not increase significantly the stellar mass content of discs; this
raise might well be within the errors. Nevertheless, we have seen that some galaxies do have very conspicuous
spiral arms, which might contain a non-negligible fraction of the stellar mass of the galaxy. Thus,
two related questions, which are relevant to studies on galactic structure and star formation,
and whose answers are not clear, emerge from this discussion:
{\bf (i)}, what is the fraction of the stellar mass of the disc of a galaxy with prominent spiral arms that resides
within the arms, and {\bf (ii)}, what is the fraction of disc galaxies that host such prominent spiral arms.
With reliable answers to these questions one would be able to properly include spiral arms as another separate
constituent of the stellar mass budget.

It was shown, in Sect. \ref{sec:noagn} (Fig. \ref{noagn}),
that not taking into account the contribution from bright AGN to the
light distribution in the central region of a galaxy can lead to severe errors in the bulge parameters. In
particular, the S\'ersic index of the bulge can be significantly overestimated\footnote{Given enough spatial
resolution, as in HST images of nearby galaxies, other central components, like nuclear star clusters,
can induce similar effects.}. Thus, one could, in principle, devise
a methodology to identify AGN using only imaging data, through these effects. The main advantage over
current methods, which use spectroscopy data, would be that imaging usually requires much less telescope
time and/or smaller telescopes. One could fit the galaxy image (aiming to find AGN, it could be that fitting
only the surface brightness profile is enough) with and without the AGN in the model. If the S\'ersic
index of the bulge, estimated in the latter fit, is larger than the one obtained in the former, over the
uncertainties (which are usually $\pm1$ for this parameter), this would be an indication of the presence
of an AGN. However, we saw that only bright, type 1 AGN produce such effects. Furthermore, there
could be large uncertainties due to the degeneracy in the possible solutions. For instance, a classical
bulge with $n=3$ could be misidentified as a pseudo-bulge, with $n=1$, containing an AGN.
In addition, as seen in Sect. \ref{sec:hzbaragn} (Table \ref{noagnt} and Fig. \ref{hzgraph}),
if the resolution is not sufficient,
the effects of the extra light from AGN are negligible. Not surprisingly, as opposed to bars, the
AGN contribution is completely smeared out by the PSF in the low resolution regime.
Such a methodology would then be severely prone to errors.

Notably, the effects of ignoring bars and AGN, and the effects of having images with low resolution,
all affect bulges more substantially than discs. Partly, this might be due to the fact that a simpler
function, an exponential, is used to describe the disc luminosity profile,
as compared to the S\'ersic function, used to describe
the bulge luminosity profile. A simple exponential function, however, might not be, in many cases,
the best choice to fit discs. The results for NGC 4608 and NGC 5701 in Sect. \ref{4608} are clear
instances, even if, perhaps, extreme cases. In, e.g., \citet[and references therein]{ebp05}, one finds
a number of cases in which the luminosity profile of the disc can be better reproduced by a
double exponential function. It is worth noting that in {\sc budda} v2.1 it is possible to use such a
function to fit discs\footnote{See http://www.mpa-garching.mpg.de/$\sim$dimitri/budda.html .}.
Nevertheless, discs are the more extended luminous galactic component and this might also
partially explain the robustness of the disc fits. The results obtained in this study suggest that,
in general, the structural parameters of discs are those which are more reliably determined.

\section{Summary and Conclusions}

I presented the results of image fitting to a sample of 17 nearby galaxies, imaged in the V and R
broadbands, including the structural parameters of bulges, discs and bars. The light from bright
AGN was taken into account when needed. A number of tests is performed to verify the reliability
of such techniques when bars and AGN are not included in the models, and when the images have a
relatively poor physical spatial resolution, which is usually the case of studies on large
samples of more distant galaxies. The main results from this work can be summarised as follows:

\begin{enumerate}

\item The ellipticity of bars, when measured as the peak in the ellipticity profile from ellipse
fits to the galaxy image, is underestimated, on average, by $\approx20\%$. To obtain the true bar
ellipticity, the contribution from the axisymmetric components (most importantly the disc)
to the galaxy image has to be considered.

\item Modelling of galaxy images is a reliable tool to determine the structural parameters of bulges,
discs and bars, even in a low resolution regime, i.e., at least up to the point where the physical
spatial resolution in the image is 1.5 Kpc, but the bulge parameters are only trustworthy if its
effective radius is not small compared to the PSF radius. The disc parameters are the most robust,
in particular the disc central surface brightness. The luminosity fractions of bulges, discs and
bars are also recovered very reliably. The bulge-to-total fraction, however, has to be corrected
for a systematic effect, using Eq. (5). The bulge-to-total fraction is a favoured parameter to be used
for quantitative morphological classification of galaxies, as opposed to the bulge S\'ersic index, since
the latter has large uncertainties compared to its usual dynamic range.

\item If bars are not modelled, when fitting barred galaxies, the structural parameters of bulges
and discs can be severely compromised, particularly the bulge effective radius.
Furthermore, the disc-to-total luminosity ratio is overestimated, on average,
by $\approx10\%$, and the bulge-to-total luminosity ratio is overestimated, on average, by
$\approx50\%$. These effects are still significant in the low resolution regime, albeit with a
reduced impact, in this case, for weaker bars.

\item If the light from bright, type 1 AGN is not modelled, when fitting their hosts, the structural
parameters of bulges can be severely compromised, particularly the bulge S\'ersic index. The
bulge-to-total luminosity ratio can be overestimated by a factor of two. However, in the low
resolution regime, the AGN contribution is smeared out by the PSF and these effects are absent.

\item Using the results concerning the biases in the estimation of the bulge and disc luminosity fractions,
due to low spatial resolution and the non-inclusion of bars in the photometric models, it is possible
to correct the stellar mass budget in the local universe, as found in the literature, to take into account
the mass in stars that reside in bars. The results are as follows. The stellar content in classical bulges
and discs is found to be $\approx13.5\%$ and $\approx58.5\%$, respectively. And the stellar content in bars is
$\approx12\%$. Nonetheless, these are rough estimates and need to be confirmed by further studies,
in particular, by the direct inclusion of bars in the models used to fit galaxy images in large samples.

\end{enumerate}

\section*{Acknowledgments}
It is a pleasure to thank Guinevere Kauffmann and Lia Athanassoula for useful discussions and for comments on a first
draft of this paper, which helped to significantly improve it, and Ronaldo de Souza, for his help in
the making of the new {\sc budda} version. I would also like to thank an anonymous referee, who contributed
with insightful suggestions.
DAG is supported by the Deutsche Forschungsgemeinschaft priority program 1177 (``Witnesses of Cosmic
History: Formation and evolution of galaxies, black holes and their environment''), and the Max Planck
Society.

\label{lastpage}

\clearpage

\begin{table*}
\centering
\begin{minipage}{111mm}
\caption{Basic Data for the Galaxies in the Sample.}
\label{basic}
\begin{tabular}{@{}llcccccl@{}}
\hline \hline
Galaxy & Type & D$_{25}$ & log R$_{25}$ & m$_{\rm B}$ & cz & d & AGN \\
(1) & (2) & (3) & (4) & (5) & (6) & (7) & (8) \\
\hline
IC 0486     & SBa            & 0.93         & 0.11         & 14.60          &     7792 & 111.3    & Sey1            \\
NGC 2110    & SAB0           & 1.70          & 0.13         & \omit           &    2064  & 29.5    & Sey2            \\
NGC 2911    & SA0(s)         & 4.07          & 0.11         & 12.21          &    3195   & 45.6     & Sey/LINER     \\
NGC 3227    & SABa(s)        & 5.37          & 0.17         & 11.59          &    1235    & 17.6     & Sey1.5         \\
NGC 4151    & SABab(rs)      & 6.31          & 0.15         & 10.90          &     1190    & 17.0     & Sey1.5          \\
NGC 4267    & SB0(s)         & 3.24          & 0.03         & 11.73          &    1123   & 16.0     & \omit            \\
NGC 4303    & SABbc(rs)      & 6.46          & 0.05         & 10.21          &    1620     & 23.1     & Sey2            \\
NGC 4314    & SBa(rs)        & 4.17          & 0.05         & 11.22          &  1146 & 16.4      & LINER            \\
NGC 4394    & SBb(r)         & 3.63          & 0.05         & 11.53  &     1036 & 14.8      & LINER            \\
NGC 4477    & SB0(s)         & 3.80           & 0.04         & 11.27          &    1441  & 20.6     & Sey2            \\
NGC 4579    & SABb(rs)       & 5.89          & 0.10         & 10.68          &   1607 & 23.0     & LINER/Sey1.9 \\
NGC 4593    & SBb(rs)        & 3.89           & 0.13         & 11.67          &    2498  & 35.7     & Sey1            \\
NGC 4608    & SB0(r)         & 3.24          & 0.08         & 11.96   &    1893 & 27.0     & \omit            \\
NGC 4665    & SB0/a(s)       & 3.80          & 0.08         & 11.50          &     872    & 12.5      & \omit             \\
NGC 5383    & SBb(rs)        & 3.16          & 0.07         & 12.18      &    2472 & 35.3     & \omit             \\
NGC 5701    & SB0/a(rs)      & 4.26           & 0.02         & 11.82    &    1601 & 22.9     & LINER             \\
NGC 5850    & SBb(r)         & 4.26           & 0.06         & 12.04  &    2637 & 37.7     & \omit            \\
\hline
\end{tabular}
Columns (1) and (2) give, respectively, the galaxy designation and morphological type,
while column (3) shows its diameter in arcminutes at the 25 B magnitude isophotal level, and
column (4) shows the decimal logarithm of its major to minor axes ratio at the same level. Columns (5)
and (6) show, respectively, the apparent B magnitude and the radial velocity in Km/s. All these data were
taken from \citet[hereafter RC3]{dev91}, except the radial velocity, taken from the Lyon Extragalactic Data
Archive (hereafter LEDA), corrected for infall of the Local Group towards Virgo. Column (7) gives the distance
to the galaxy in Mpc, using the radial velocity in column (6) and H$_0=70$ Km s$^{-1}$ Mpc$^{-1}$.
Column (8) presents the AGN classification according
to the NASA Extragalactic Database (hereafter NED).
\end{minipage}
\end{table*}

\begin{table*}
\centering
\begin{minipage}{140mm}
\caption{Galaxy Structural Parameters in the R-band.}
\label{rres}
\begin{tabular}{@{}lcccccccccccc@{}}
\hline \hline
Galaxy & $\mu_0$ & $h$ & $\mu_e$ & $r_e$ & $n$ & $n_{\rm Bar}$ & $L_{\rm Bar}$ & $\epsilon_{\rm Bar}$ & $c$ & B/T & D/T & Bar/T \\
(1) & (2) & (3) & (4) & (5) & (6) & (7) & (8) & (9) & (10) & (11) & (12) & (13) \\
\hline
IC 486   & 19.7 &   8.9 & 17.6 &  1.1 & 2.1 & 0.49 & 12.7 & 0.54 & 2.12 & 0.213 & 0.579 & 0.208 \\
NGC 2110 & 18.9 &  21.8 & 18.2 &  6.8 & 2.7 & \omit& \omit& \omit& \omit& 0.390 & 0.610 & \omit \\
NGC 2911 & 20.9 &  41.0 & 19.6 &  7.1 & 3.0 & \omit& \omit& \omit& \omit& 0.354 & 0.646 & \omit \\
NGC 3227 & 19.1 &  38.3 & 18.2 &  3.7 & 1.1 & 1.00 & 45.8 & 0.74 & 2.80 & 0.068 & 0.876 & 0.017 \\
NGC 4151$^*$ & 20.2 &  33.7 & 18.0 &  4.7 & 3.0 & 0.60 & 82.1 & 0.60 & 3.07 & 0.327 & 0.503 & 0.082 \\
NGC 4267 & 20.2 &  31.1 & 18.2 &  4.8 & 3.1 & 0.77 & 25.5 & 0.60 & 2.42 & 0.356 & 0.593 & 0.051 \\
NGC 4303 & 19.6 &  41.7 & 17.6 &  3.0 & 1.0 & 0.67 & 33.3 & 0.65 & 2.80 & 0.066 & 0.898 & 0.028 \\
NGC 4314 & 21.3 &  53.0 & 19.2 & 10.2 & 2.2 & 0.40 & 95.4 & 0.75 & 2.89 & 0.296 & 0.397 & 0.308 \\
NGC 4394 & 20.4 &  37.4 & 18.2 &  4.2 & 1.8 & 0.56 & 53.1 & 0.70 & 2.76 & 0.186 & 0.676 & 0.138 \\
NGC 4477 & 19.5 &  28.7 & 18.0 &  5.0 & 2.0 & 0.66 & 32.9 & 0.50 & 1.96 & 0.183 & 0.689 & 0.128 \\
NGC 4579 & 19.5 &  39.5 & 17.9 &  5.2 & 1.4 & 0.38 & 37.5 & 0.50 & 2.04 & 0.127 & 0.749 & 0.115 \\
NGC 4593 & 20.3 &  43.0 & 18.7 &  6.5 & 0.9 & 0.66 & 62.2 & 0.73 & 2.72 & 0.157 & 0.671 & 0.127 \\
NGC 4608-I  & 20.7 &  40.7 & 18.3 &  5.2 & 1.7 & 0.58 & 57.6 & 0.66 & 2.02 & 0.257 & 0.565 & 0.178 \\
NGC 4608-II & 20.2 &  33.4 & 18.7 &  6.9 & 2.1 & 0.58 & 57.6 & 0.66 & 2.02 & 0.327 & 0.491 & 0.182 \\
NGC 4665$^*$ & 21.5 &  63.6 & 19.4 &  5.9 & 2.0 & 1.06 & 54.0 & 0.65 & 1.99 & 0.152 & 0.676 & 0.172 \\
NGC 5383 & 20.5 &  28.1 & 19.5 &  7.0 & 0.9 & 0.31 & 62.0 & 0.69 & 2.96 & 0.166 & 0.651 & 0.183 \\
NGC 5701 & 21.2 &  42.9 & 19.2 &  6.5 & 3.2 & 0.62 & 44.7 & 0.56 & 2.35 & 0.278 & 0.501 & 0.221 \\
NGC 5850$^*$ & 22.0 & 116.6 & 19.0 &  5.3 & 2.1 & 0.64 & 60.9 & 0.61 & 2.44 & 0.172 & 0.681 & 0.147 \\
\hline
\end{tabular}
Structural parameters of bulges, discs and bars. Column (1) gives the galaxy name, while columns (2) and (3)
show, respectively, the disc central surface brightness and scale-length. Columns (4), (5) and (6) show the
bulge effective surface brightness, effective radius and S\'ersic index, respectively. Columns (7) and (8)
show the S\'ersic index of the bar luminosity profile and the length of the bar semi-major axis, respectively.
Column (9) shows the bar ellipticity, whereas column (10) shows the shape index of the bar isophotes. Finally,
columns (11), (12) and (13) give, respectively, the estimated luminosity fractions of bulge, disc and bar.
Luminosity parameters are in units of mag arcsec$^{-2}$ and scale-lengths in arcseconds. Galaxies marked with
$^*$ have uncertain estimates for the disc parameters. The two rows for NGC 4608 correspond to the fits with
a type I and a type II disc, as indicated.
\end{minipage}
\end{table*}

\begin{table*}
\centering
\begin{minipage}{138mm}
\caption{Galaxy Structural Parameters in the V-band.}
\label{vres}
\begin{tabular}{@{}lcccccccccccc@{}}
\hline \hline
Galaxy & $\mu_0$ & $h$ & $\mu_e$ & $r_e$ & $n$ & $n_{\rm Bar}$ & $L_{\rm Bar}$ & $\epsilon_{\rm Bar}$ & $c$ & B/T & D/T & Bar/T \\
(1) & (2) & (3) & (4) & (5) & (6) & (7) & (8) & (9) & (10) & (11) & (12) & (13) \\
\hline
IC 486   & 20.1 &   8.3 & 18.0 &  1.1 & 2.7 & 0.47 & 12.6 & 0.57 & 2.24 & 0.239 & 0.562 & 0.199 \\
NGC 2110 & 18.8 &  17.0 & 18.6 &  6.1 & 3.4 & \omit& \omit& \omit& \omit& 0.348 & 0.652 & \omit \\
NGC 2911 & 21.4 &  32.7 & 20.2 &  7.0 & 2.6 & \omit& \omit& \omit& \omit& 0.389 & 0.611 & \omit \\
NGC 3227 & 19.5 &  39.9 & 19.2 &  4.5 & 0.6 & 1.00 & 45.8 & 0.74 & 2.80 & 0.049 & 0.905 & 0.007 \\
NGC 4151$^*$ & 20.6 &  33.7 & 18.2 &  4.2 & 3.8 & 0.60 & 82.1 & 0.60 & 3.07 & 0.361 & 0.481 & 0.082 \\
NGC 4267 & 21.0 &  33.6 & 18.8 &  5.3 & 3.4 & 0.77 & 22.3 & 0.60 & 2.42 & 0.396 & 0.557 & 0.048 \\
NGC 4303 & 20.1 &  43.7 & 18.2 &  3.1 & 1.0 & 0.67 & 33.3 & 0.65 & 2.80 & 0.061 & 0.904 & 0.024 \\
NGC 4314 & 22.0 &  58.4 & 19.8 & 10.9 & 2.0 & 0.37 & 95.7 & 0.75 & 2.89 & 0.304 & 0.395 & 0.301 \\
NGC 4394 & 20.8 &  37.0 & 18.7 &  4.2 & 1.9 & 0.60 & 53.0 & 0.71 & 2.76 & 0.184 & 0.685 & 0.131 \\
NGC 4477 & 19.9 &  27.5 & 18.6 &  4.9 & 1.8 & 0.71 & 32.9 & 0.50 & 2.05 & 0.168 & 0.695 & 0.137 \\
NGC 4579 & 19.9 &  38.3 & 18.4 &  5.4 & 1.2 & 0.29 & 38.5 & 0.52 & 1.96 & 0.125 & 0.754 & 0.109 \\
NGC 4593 & 20.7 &  46.2 & 19.3 &  6.9 & 0.7 & 0.63 & 62.2 & 0.73 & 2.67 & 0.133 & 0.710 & 0.112 \\
NGC 4608 & 21.5 &  46.8 & 18.9 &  5.5 & 1.9 & 0.63 & 57.8 & 0.63 & 2.02 & 0.268 & 0.540 & 0.192 \\
NGC 4665$^*$ & 20.7 &  48.7 & 19.2 &  6.4 & 2.1 & 1.06 & 54.0 & 0.65 & 1.99 & 0.149 & 0.687 & 0.164 \\
NGC 5383 & 21.0 &  30.5 & 19.9 &  7.1 & 0.8 & 0.38 & 62.0 & 0.70 & 2.89 & 0.147 & 0.698 & 0.155 \\
NGC 5701 & 22.0 &  63.2 & 19.9 &  7.0 & 3.2 & 0.62 & 45.6 & 0.56 & 2.35 & 0.240 & 0.567 & 0.193 \\
NGC 5850$^*$ & 22.4 & 118.9 & 19.5 &  5.1 & 2.0 & 0.58 & 60.9 & 0.62 & 1.99 & 0.152 & 0.719 & 0.130 \\
\hline
\end{tabular}
Structural parameters of bulges, discs and bars. As in Table 1 but from the V-band images. Luminosity parameters
are in units of mag arcsec$^{-2}$ and scale-lengths in arcseconds. Galaxies marked with $^*$ have uncertain estimates
for the disc parameters.
\end{minipage}
\end{table*}

\begin{table*}
\centering
\begin{minipage}{58mm}
\caption{Linear regression results for Fig. \ref{hz}.}
\label{regress}
\begin{tabular}{@{}lccc@{}}
\hline \hline
Parameter & cc & $a$ & $b$ \\
\hline
$\mu_0$     &   0.97   &   0.91$\pm$0.07   &   1.62$\pm$1.43     \\
$h$         &   0.91   &   0.77$\pm$0.10   &   4.95$\pm$3.63     \\
$\mu_e$     &   0.79   &   0.63$\pm$0.14   &   6.59$\pm$2.62     \\
$r_e^*$     &   0.58   &   0.48$\pm$0.19   &   3.21$\pm$1.14     \\
$r_e^\star$ &   0.79   &   0.59$\pm$0.17   &   2.02$\pm$1.15     \\
$n^*$       &   0.51   &   0.33$\pm$0.16   &   0.83$\pm$0.34     \\
$n^\star$   &   0.92   &   0.65$\pm$0.11   &   0.19$\pm$0.23     \\
B/T$^*$     &   0.91   &   0.86$\pm$0.11   &   0.10$\pm$0.03     \\
B/T$^\star$ &   0.93   &   0.89$\pm$0.13   &   0.08$\pm$0.03     \\
D/T         &   0.97   &   0.93$\pm$0.07   &   0.00$\pm$0.05     \\
Bar/T       &   0.98   &   0.96$\pm$0.06   &  -0.02$\pm$0.01     \\
\hline
\end{tabular}
Parameters obtained from the linear regressions in Fig. \ref{hz}: cc is the correlation coefficient, $a$ is the slope
of the line and $b$ its intercept. Thus, the fitted line to, e.g., $\mu_0$, can be written as
$y=0.91(\pm0.07)x+1.62(\pm1.43)$. Uncertainties quoted are 1 $\sigma$ errors from the fit. Luminosity parameters
are in units of mag arcsec$^{-2}$ and scale-lengths in arcseconds. For $r_e$, $n$ and B/T,
parameters with $^*$ appended correspond to fits using all data points, whereas those with
$^\star$ appended correspond to fits where the data points deemed unreliable were excluded.
\end{minipage}
\end{table*}

\begin{table*}
\centering
\begin{minipage}{177mm}
\caption{Bar effects on $r_e$ and B/T at different resolutions.}
\label{nobart}
\begin{tabular}{@{}lccccccccc@{}}
\hline \hline
\omit & \omit & $z\approx0.005$ & \omit & \omit & \omit & $z=0.05$ & \omit & \omit & \omit \\
Galaxy & Bar/T & $r_e$ (Bar) & $r_e$ (no Bar) & B/T (Bar) & B/T (no Bar) &
$r_e$ (Bar) & $r_e$ (no Bar) & B/T (Bar) & B/T (no Bar) \\
(1) & (2) & (3) & (4) & (5) & (6) & (7) & (8) & (9) & (10) \\
\hline
NGC 4314 & 0.308 & 10.9 & 16.6 (52\%) & 0.296 & 0.591 (100\%) & 6.8 & 14.8 (118\%) & 0.320 & 0.618 (93\%) \\
NGC 4394 & 0.138 & 4.2  & 5.3  (26\%) & 0.186 & 0.253 (36\%)  & 7.2 & 10.5 (46\%)  & 0.309 & 0.375 (21\%) \\
NGC 4477 & 0.128 & 4.9  & 7.1  (45\%) & 0.183 & 0.308 (68\%)  & 6.9 & 8.2  (19\%)  & 0.291 & 0.370 (27\%) \\
\hline
\end{tabular}
Column (1) gives the galaxy name and column (2) shows the estimated bar luminosity fraction.
Columns (3) and (4) show the effective radius of the bulge, $r_e$, with and without a bar in the model,
respectively. Similarly, columns (5) and (6) show the estimated bulge luminosity fraction,
with and without a bar in the model. The latter five columns refer to the original galaxy images.
Columns (7) to (10) are similar to columns (3) to (6) but refer to the artificially redshifted images.
The effective radius is in arcseconds and, for the redshifted images, scaled back to the original galaxy distance.
The values in parenthesis give the relative change in the parameter when omitting the bar.
\end{minipage}
\end{table*}

\begin{table*}
\centering
\begin{minipage}{129mm}
\caption{AGN effects on the bulge S\'ersic index at different resolutions.}
\label{noagnt}
\begin{tabular}{@{}lccccc@{}}
\hline \hline
\omit & \omit & $z\approx0.005$ & \omit & $z=0.05$ & \omit \\
Galaxy & AGN/T & $n$ (with AGN) & $n$ (without AGN) & $n$ (with AGN) & $n$ (without AGN) \\
(1) & (2) & (3) & (4) & (5) & (6) \\
\hline
NGC 4303 & 0.008 & 1.0 & 1.1 & 1.4 & 1.5 \\
NGC 4579 & 0.009 & 1.4 & 2.6 & 0.9 & 0.8 \\
NGC 4593 & 0.045 & 0.9 & 4.3 & 1.1 & 0.9 \\
\hline
\end{tabular}
Column (1) gives the galaxy name, while column (2) shows the estimated AGN luminosity fraction. Column (3)
shows the bulge S\'ersic index when the AGN is included in the model, while column (4) shows the same
parameter when the AGN is not taken into account. The latter three columns refer to the original galaxy images.
Columns (5) and (6) are similar to columns (3) and (4), respectively, but refer to the artificially redshifted
images.
\end{minipage}
\end{table*}

\clearpage

\begin{table*}
\vskip 3cm
\centering
\Huge{For the complete Figure 1, please download PDF file at
http://www.mpa-garching.mpg.de/$\sim$dimitri/solo$\_$rev.pdf .}
\end{table*}

\clearpage

\clearpage

\clearpage

\clearpage

\begin{thebibliography}{}
  \bibitem[Allen et al.(2006)]{all06} Allen, P. D., Driver, S. P., Graham, A. W., Cameron, E., Liske,
J., \& de Propis, R. 2006, MNRAS, 371, 2
  \bibitem[Athanassoula(1992)]{ath92} Athanassoula, E. 1992, MNRAS, 259, 328
  \bibitem[Athanassoula(2002)]{ath02b} Athanassoula, E. 2002, ApJL, 569, 83
  \bibitem[Athanassoula(2003)]{ath03} Athanassoula, E. 2003, MNRAS, 341, 1179
  \bibitem[Athanassoula \& Misiriotis(2002)]{ath02} Athanassoula, E., \&
Misiriotis, A. 2002, MNRAS, 330, 35
  \bibitem[Athanassoula et al.(1990)]{ath90} Athanassoula, E., et al. 1990, MNRAS, 245, 130
  \bibitem[Binney \& Tremaine(1987)]{bin87} Binney, J., \& Tremaine, S. 1987, in Galactic Dynamics, Princeton
University Press
  \bibitem[Buta et al.(2006)]{but06} Buta, R., Laurikainen, E., Salo,
H, Block, D. L., \& Knapen, J. H. 2006, AJ, 132, 1859
  \bibitem[Caon, Capaccioli \& D'Onofrio(1993)]{cao93} Caon, N., Capaccioli, M., \& D'Onofrio, M. 1993,
MNRAS, 265, 1013
  \bibitem[de Jong(1995)]{dej95} de Jong, R. S. 1995, PhD thesis, University of Groningen, The Netherlands
  \bibitem[de Jong(1996)]{dej96} de Jong, R. S. 1996, A\&A, 313, 45
  \bibitem[de Souza, Gadotti \& dos Anjos(2004)]{des04} de Souza, R. E., Gadotti, D. A., \& dos Anjos, S. 2004, ApJS,
153, 411
  \bibitem[de Vaucouleurs et al.(1991)]{dev91} de Vaucouleurs, G., de Vaucouleurs, A., Corwin, H. G.,
Buta, R. J., Paturel, G., \& Fouque, P. 1991, Third Reference Catalog of Bright Galaxies (New York:
Springer-Verlag) {\bf (RC3)}
  \bibitem[D'Onofrio(2001)]{don01} D'Onofrio, M. 2001, MNRAS, 326, 1517
  \bibitem[Driver et al.(2007)]{dri07} Driver, S. P., Allen, P. D., Liske, J., \& Graham, A. W. 2007, ApJL, in press
(astro-ph/0701728)
  \bibitem[Elmegreen \& Elmegreen(1985)]{elm85} Elmegreen, B. G., \& Elmegreen, D. M. 1985, ApJ, 288, 438
  \bibitem[Erwin(2005)]{erw05} Erwin, P. 2005, MNRAS, 364, 283
  \bibitem[Erwin, Beckman \& Pohlen(2005)]{ebp05} Erwin, P., Beckman, J. E., \& Pohlen, M. 2005, ApJL, 626, L81
  \bibitem[Eskridge et al.(2000)]{esk00} Eskridge, P. B., et al. 2000, AJ, 119, 536
  \bibitem[Freeman(1970)]{fre70} Freeman, K. C., ApJ, 160, 811
  \bibitem[Gadotti et al.(2007)]{gad07} Gadotti, D. A., Athanassoula, E., Carrasco, L., Bosma, A., de Souza, R.,
Recillas, E. 2007, MNRAS, 381, 943
  \bibitem[Gadotti \& de Souza(2003)]{gad03} Gadotti, D. A., \& de Souza, R. E. 2003, ApJL, 583, L75
  \bibitem[Gadotti \& de Souza(2005)]{gad05} Gadotti, D. A., \& de Souza, R. E.
2005, ApJ, 629, 797
  \bibitem[Gadotti \& de Souza(2006)]{gad06} Gadotti, D. A., \& de Souza, R. E.
2006, ApJS, 163, 270 {\bf (GdS06)}
  \bibitem[Gadotti \& dos Anjos(2001)]{gad01} Gadotti, D. A., \& dos Anjos, S.
2001, AJ, 122, 1298
  \bibitem[Gadotti \& Kauffmann(2007)]{gadlp} Gadotti, D. A., \& Kauffmann, G. 2007, in: Stellar Populations
as Building Blocks of Galaxies, IAU Symp. 241, A. Vazdekis, R. Peletier (eds.), 507
  \bibitem[Genzel et al.(2006)]{gen06} Genzel, R., et al. 2006, Nature, 442, 786
  \bibitem[H\"aussler et al.(2007)]{hau07} H\"aussler, B., et al. 2007, ApJS, 172, 615
  \bibitem[Hubble(1926)]{hub26} Hubble, E. P. 1926, ApJ, 64, 321
  \bibitem[Hubble(1936)]{hub36} Hubble, E. P. 1936, in: The Realm of the Nebulae (New Haven: Yale University
Press)
  \bibitem[Huertas-Company et al.(2007)]{hue07} Huertas-Company, M., Rouan, D., Soucail, G., Le F\`evre, O.,
Tasca, L., Contini, T. 2007, A\&A, 468, 937
  \bibitem[Khosroshahi, Wadadekar \& Kembhavi(2000)]{kho00} Khosroshahi, H. G., Wadadekar, Y, \& Kembhavi, A.
2000, ApJ, 533, 162
  \bibitem[Koekemoer et al.(2007)]{koe07} Koekemoer, A. M., et al. 2007, ApJS, 172, 196
  \bibitem[Kormendy(1977)]{kor77} Kormendy, J. 1977, ApJ, 218, 333
  \bibitem[Kormendy \& Kennicutt(2004)]{kor04} Kormendy, J., \& Kennicutt, R. C. 2004,
ARA\&A, 42, 603
  \bibitem[Laurikainen et al.(2004)]{lau04} Laurikainen, E., Salo, H., Buta, R., \& Vasylyev, S. 2004, MNRAS,
355, 1251
  \bibitem[Laurikainen, Salo \& Buta(2005)]{lau05} Laurikainen, E., Salo, H., \& Buta, R. 2005, MNRAS,
362, 1319
  \bibitem[Laurikainen et al.(2006)]{lau06} Laurikainen, E., Salo, H., Buta, R., Knapen, J., Speltincx, T., \&
Block, D. 2006, AJ, 132, 2634
  \bibitem[Marinova \& Jogee(2007)]{mar07} Marinova, I., \& Jogee, S. 2007, ApJ, 659, 1176
  \bibitem[Marleau \& Simard(1998)]{mar98} Marleau, F. R. \& Simard, L. 1998, ApJ, 507, 585
  \bibitem[Martinez-Valpuesta, Knapen \& Buta(2007)]{val07} Martinez-Valpuesta, I., Knapen, J. H., \& Buta, R.
2007, AJ, 134, 1863
  \bibitem[M\"ollenhoff \& Heidt(2001)]{mol01} M\"ollenhoff, C. \& Heidt, J. 2001, A\&A, 368, 16
  \bibitem[Moffat(1969)]{mof69} Moffat, A. F. J. 1969, A\&A, 3, 455
  \bibitem[Patsis, Skokos \& Athanassoula(2003)]{pat03} Patsis, P. A., Skokos, Ch., Athanassoula, E.
2003, MNRAS, 342, 69
  \bibitem[Peng et al.(2002)]{pen02} Peng, C. Y., Ho, L. C., Impey, C. D., \& Rix, H-W. 2002, AJ, 124, 266
  \bibitem[Pignatelli, Fasano \& Cassata(2006)]{pig06} Pignatelli, E., Fasano, G., \& Cassata, P. 2006, A\&A, 446,
373
  \bibitem[Reese et al.(2007)]{ree07} Reese, A. S., Williams, T. B., Sellwood, J. A., Barnes, E. I., \& Powell,
B. A. 2007, AJ, 133, 2846
  \bibitem[Sellwood \& Wilkinson(1993)]{sel93} Sellwood, J. A. \& Wilkinson, A. 1993, Rep. Pr. Phys., 56, 173
  \bibitem[S\'ersic(1968)]{ser68} S\'ersic, J. L. 1968, Atlas de Galaxias Australes, Observatorio Astronomico, Cordoba
  \bibitem[Tasca \& White(2005)]{tas05} Tasca, L. A. M., \& White, S. D. M. 2005, MNRAS, submitted (astro-ph/0507249)
  \bibitem[Trujillo et al.(2001a)]{tru01a} Trujillo, I., Aguerri, J. A. L., Cepa, J., Guti\'errez, C. M. 2001, MNRAS,
321, 269
  \bibitem[Trujillo et al.(2001b)]{tru01b} Trujillo, I., Aguerri, J. A. L., Cepa, J., Guti\'errez, C. M. 2001, MNRAS,
328, 977

\end{thebibliography}
\end{document}